\documentclass[twocolumn]{aastex631}
\usepackage{CJK}
\shorttitle{DES Jupiter Trojans}
\shortauthors{Pan et al.}
\graphicspath{{./}{figures/}}
\begin{document}
\begin{CJK*}{UTF8}{gbsn}
\reportnum{DES-2022-0701}
\reportnum{FERMILAB-PUB-22-628}
\title{Photometric Properties of Jupiter Trojans detected by the Dark Energy Survey}

\author[0000-0001-9685-5756]{Jiaming Pan (潘嘉明)}
\affiliation{Department of Astronomy, University of Michigan \\
Ann Arbor, 48109 MI, USA}
\affiliation{Department of Physics, University of Michigan\\ Ann Arbor, MI 48109, USA\\}

\author[0000-0001-7737-6784]{Hsing Wen Lin (林省文)}
\affiliation{Department of Physics, University of Michigan\\ Ann Arbor, MI 48109, USA\\}

\author[0000-0001-6942-2736]{David W. Gerdes}
\affiliation{Department of Astronomy, University of Michigan \\
Ann Arbor, 48109 MI, USA}
\affiliation{Department of Physics, University of Michigan\\ Ann Arbor, MI 48109, USA\\}

\author[0000-0003-4827-5049]{Kevin J. Napier}
\affiliation{Department of Physics, University of Michigan\\ Ann Arbor, MI 48109, USA\\}

\author[0000-0002-3102-9442]{Jichi Wang (王骥驰)}
\affiliation{Department of Physics, University of Michigan\\ Ann Arbor, MI 48109, USA\\}

% Author list file generated with: mkauthlist.py 1.3.0+14.gcc6daf1 
% mkauthlist.py -sb -j aastex61 --orcid DES-2022-0701_author_list.csv 

%%% DES Builders
\author{T.~M.~C.~Abbott}
\affiliation{Cerro Tololo Inter-American Observatory, NSF's National Optical-Infrared Astronomy Research Laboratory, Casilla 603, La Serena, Chile}

\author{M.~Aguena}
\affiliation{Laborat\'orio Interinstitucional de e-Astronomia - LIneA, Rua Gal. Jos\'e Cristino 77, Rio de Janeiro, RJ - 20921-400, Brazil}

\author[0000-0002-7069-7857]{S.~Allam}
\altaffiliation{Deceased.}
\affiliation{Fermi National Accelerator Laboratory, P. O. Box 500, Batavia, IL 60510, USA}

\author{O.~Alves}
\affiliation{Department of Physics, University of Michigan\\ Ann Arbor, MI 48109, USA\\}
\affiliation{Laborat\'orio Interinstitucional de e-Astronomia - LIneA, Rua Gal. Jos\'e Cristino 77, Rio de Janeiro, RJ - 20921-400, Brazil}

\author{D.~Bacon}
\affiliation{Institute of Cosmology and Gravitation, University of Portsmouth, Portsmouth, PO1 3FX, UK}

\author[0000-0003-0743-9422]{P.~H.~Bernardinelli}
\affiliation{Astronomy Department, University of Washington, Box 351580, Seattle, WA 98195, USA}

\author[0000-0002-8613-8259]{G.~M.~Bernstein}
\affiliation{Department of Physics and Astronomy, University of Pennsylvania, Philadelphia, PA 19104, USA}

\author{E.~Bertin}
\affiliation{CNRS, UMR 7095, Institut d'Astrophysique de Paris, F-75014, Paris, France}
\affiliation{Sorbonne Universit\'es, UPMC Univ Paris 06, UMR 7095, Institut d'Astrophysique de Paris, F-75014, Paris, France}

\author[0000-0002-8458-5047]{D.~Brooks}
\affiliation{Department of Physics \& Astronomy, University College London, Gower Street, London, WC1E 6BT, UK}

\author{D.~L.~Burke}
\affiliation{Kavli Institute for Particle Astrophysics \& Cosmology, P. O. Box 2450, Stanford University, Stanford, CA 94305, USA}
\affiliation{SLAC National Accelerator Laboratory, Menlo Park, CA 94025, USA}

\author[0000-0003-3044-5150]{A.~Carnero~Rosell}
\affiliation{Instituto de Astrofisica de Canarias, E-38205 La Laguna, Tenerife, Spain}
\affiliation{Laborat\'orio Interinstitucional de e-Astronomia - LIneA, Rua Gal. Jos\'e Cristino 77, Rio de Janeiro, RJ - 20921-400, Brazil}
\affiliation{Universidad de La Laguna, Dpto. Astrofísica, E-38206 La Laguna, Tenerife, Spain}

\author[0000-0002-4802-3194]{M.~Carrasco~Kind}
\affiliation{Center for Astrophysical Surveys, National Center for Supercomputing Applications, 1205 West Clark St., Urbana, IL 61801, USA}
\affiliation{Department of Astronomy, University of Illinois at Urbana-Champaign, 1002 W. Green Street, Urbana, IL 61801, USA}

\author[0000-0002-3130-0204]{J.~Carretero}
\affiliation{Institut de F\'{\i}sica d'Altes Energies (IFAE), The Barcelona Institute of Science and Technology, Campus UAB, 08193 Bellaterra (Barcelona) Spain}

\author{M.~Costanzi}
\affiliation{Astronomy Unit, Department of Physics, University of Trieste, via Tiepolo 11, I-34131 Trieste, Italy}
\affiliation{INAF-Osservatorio Astronomico di Trieste, via G. B. Tiepolo 11, I-34143 Trieste, Italy}
\affiliation{Institute for Fundamental Physics of the Universe, Via Beirut 2, 34014 Trieste, Italy}

\author{L.~N.~da Costa}
\affiliation{Laborat\'orio Interinstitucional de e-Astronomia - LIneA, Rua Gal. Jos\'e Cristino 77, Rio de Janeiro, RJ - 20921-400, Brazil}

\author{M.~E.~S.~Pereira}
\affiliation{Hamburger Sternwarte, Universit\"{a}t Hamburg, Gojenbergsweg 112, 21029 Hamburg, Germany}

\author[0000-0001-8318-6813]{J.~De~Vicente}
\affiliation{Centro de Investigaciones Energ\'eticas, Medioambientales y Tecnol\'ogicas (CIEMAT), Madrid, Spain}

\author[0000-0002-0466-3288]{S.~Desai}
\affiliation{Department of Physics, IIT Hyderabad, Kandi, Telangana 502285, India}

\author{P.~Doel}
\affiliation{Department of Physics \& Astronomy, University College London, Gower Street, London, WC1E 6BT, UK}

\author{I.~Ferrero}
\affiliation{Institute of Theoretical Astrophysics, University of Oslo. P.O. Box 1029 Blindern, NO-0315 Oslo, Norway}

\author{D.~Friedel}
\affiliation{Center for Astrophysical Surveys, National Center for Supercomputing Applications, 1205 West Clark St., Urbana, IL 61801, USA}

\author[0000-0003-4079-3263]{J.~Frieman}
\affiliation{Fermi National Accelerator Laboratory, P. O. Box 500, Batavia, IL 60510, USA}
\affiliation{Kavli Institute for Cosmological Physics, University of Chicago, Chicago, IL 60637, USA}

\author[0000-0002-9370-8360]{J.~Garc\'ia-Bellido}
\affiliation{Instituto de Fisica Teorica UAM/CSIC, Universidad Autonoma de Madrid, 28049 Madrid, Spain}

\author{M.~Gatti}
\affiliation{Department of Physics and Astronomy, University of Pennsylvania, Philadelphia, PA 19104, USA}

\author{R.~A.~Gruendl}
\affiliation{Center for Astrophysical Surveys, National Center for Supercomputing Applications, 1205 West Clark St., Urbana, IL 61801, USA}
\affiliation{Department of Astronomy, University of Illinois at Urbana-Champaign, 1002 W. Green Street, Urbana, IL 61801, USA}

\author[0000-0003-3023-8362]{J.~Gschwend}
\affiliation{Laborat\'orio Interinstitucional de e-Astronomia - LIneA, Rua Gal. Jos\'e Cristino 77, Rio de Janeiro, RJ - 20921-400, Brazil}
\affiliation{Observat\'orio Nacional, Rua Gal. Jos\'e Cristino 77, Rio de Janeiro, RJ - 20921-400, Brazil}

\author[0000-0001-6718-2978]{K.~Herner}
\affiliation{Fermi National Accelerator Laboratory, P. O. Box 500, Batavia, IL 60510, USA}

\author{S.~R.~Hinton}
\affiliation{School of Mathematics and Physics, University of Queensland,  Brisbane, QLD 4072, Australia}

\author{D.~L.~Hollowood}
\affiliation{Santa Cruz Institute for Particle Physics, Santa Cruz, CA 95064, USA}

\author[0000-0002-6550-2023]{K.~Honscheid}
\affiliation{Center for Cosmology and Astro-Particle Physics, The Ohio State University, Columbus, OH 43210, USA}
\affiliation{Department of Physics, The Ohio State University, Columbus, OH 43210, USA}

\author[0000-0001-5160-4486]{D.~J.~James}
\affiliation{Center for Astrophysics $\vert$ Harvard \& Smithsonian, 60 Garden Street, Cambridge, MA 02138, USA}

\author[0000-0003-0120-0808]{K.~Kuehn}
\affiliation{Australian Astronomical Optics, Macquarie University, North Ryde, NSW 2113, Australia}
\affiliation{Lowell Observatory, 1400 Mars Hill Rd, Flagstaff, AZ 86001, USA}

\author[0000-0003-2511-0946]{N.~Kuropatkin}
\affiliation{Fermi National Accelerator Laboratory, P. O. Box 500, Batavia, IL 60510, USA}

\author{M.~March}
\affiliation{Department of Physics and Astronomy, University of Pennsylvania, Philadelphia, PA 19104, USA}

\author[0000-0002-1372-2534]{F.~Menanteau}
\affiliation{Center for Astrophysical Surveys, National Center for Supercomputing Applications, 1205 West Clark St., Urbana, IL 61801, USA}
\affiliation{Department of Astronomy, University of Illinois at Urbana-Champaign, 1002 W. Green Street, Urbana, IL 61801, USA}

\author[0000-0002-6610-4836]{R.~Miquel}
\affiliation{Instituci\'o Catalana de Recerca i Estudis Avan\c{c}ats, E-08010 Barcelona, Spain}
\affiliation{Institut de F\'{\i}sica d'Altes Energies (IFAE), The Barcelona Institute of Science and Technology, Campus UAB, 08193 Bellaterra (Barcelona) Spain}

\author{F.~Paz-Chinch\'{o}n}
\affiliation{Center for Astrophysical Surveys, National Center for Supercomputing Applications, 1205 West Clark St., Urbana, IL 61801, USA}
\affiliation{Institute of Astronomy, University of Cambridge, Madingley Road, Cambridge CB3 0HA, UK}

\author[0000-0001-9186-6042]{A.~Pieres}
\affiliation{Laborat\'orio Interinstitucional de e-Astronomia - LIneA, Rua Gal. Jos\'e Cristino 77, Rio de Janeiro, RJ - 20921-400, Brazil}
\affiliation{Observat\'orio Nacional, Rua Gal. Jos\'e Cristino 77, Rio de Janeiro, RJ - 20921-400, Brazil}

\author[0000-0002-2598-0514]{A.~A.~Plazas~Malag\'on}
\affiliation{Department of Astrophysical Sciences, Princeton University, Peyton Hall, Princeton, NJ 08544, USA}

\author{M.~Raveri}
\affiliation{Department of Physics and Astronomy, University of Pennsylvania, Philadelphia, PA 19104, USA}

\author{M.~Rodriguez-Monroy}
\affiliation{Centro de Investigaciones Energ\'eticas, Medioambientales y Tecnol\'ogicas (CIEMAT), Madrid, Spain}

\author[0000-0002-9328-879X]{A.~K.~Romer}
\affiliation{Department of Physics and Astronomy, Pevensey Building, University of Sussex, Brighton, BN1 9QH, UK}

\author[0000-0002-9646-8198]{E.~Sanchez}
\affiliation{Centro de Investigaciones Energ\'eticas, Medioambientales y Tecnol\'ogicas (CIEMAT), Madrid, Spain}

\author[0000-0001-9504-2059]{M.~Schubnell}
\affiliation{Department of Physics, University of Michigan\\ Ann Arbor, MI 48109, USA\\}

\author[0000-0002-1831-1953]{I.~Sevilla-Noarbe}
\affiliation{Centro de Investigaciones Energ\'eticas, Medioambientales y Tecnol\'ogicas (CIEMAT), Madrid, Spain}

\author[0000-0002-3321-1432]{M.~Smith}
\affiliation{School of Physics and Astronomy, University of Southampton,  Southampton, SO17 1BJ, UK}

\author[0000-0002-7047-9358]{E.~Suchyta}
\affiliation{Computer Science and Mathematics Division, Oak Ridge National Laboratory, Oak Ridge, TN 37831}

\author[0000-0003-1704-0781]{G.~Tarle}
\affiliation{Department of Physics, University of Michigan\\ Ann Arbor, MI 48109, USA\\}

\author[0000-0001-7211-5729]{D.~Tucker}
\affiliation{Fermi National Accelerator Laboratory, P. O. Box 500, Batavia, IL 60510, USA}

\author[0000-0002-7123-8943]{A.~R.~Walker}
\affiliation{Cerro Tololo Inter-American Observatory, NSF's National Optical-Infrared Astronomy Research Laboratory, Casilla 603, La Serena, Chile}

\author{N.~Weaverdyck}
\affiliation{Department of Physics, University of Michigan\\ Ann Arbor, MI 48109, USA\\}
\affiliation{Lawrence Berkeley National Laboratory, 1 Cyclotron Road, Berkeley, CA 94720, USA}

%\collaboration{(DES Collaboration)}

%\collaboration{6}{(AAS Journals Data Editors)}

%\author{Butler Burton}
%\affiliation{Leiden University}
%\affiliation{AAS Journals Associate Editor-in-Chief}

%\altaffiliation{AASTeX v6+ programmer}

%\affiliation{AAS Director of Publishing}
%%\affiliation{American Astronomical Society \\
%1667 K Street NW, Suite 800 \\

%\author{Magaret Donnelly}
%\affiliation{IOP Publishing, Washington, DC 20005}



\begin{abstract}

The Jupiter Trojans are a large group of asteroids that are co-orbiting with Jupiter near its L4 and L5 Lagrange points. The study of Jupiter Trojans is crucial for testing different models of planet formation that are directly related to our understanding of solar system evolution. In this work, we select known Jupiter Trojans listed by the Minor Planet Center (MPC) from the full six years dataset (Y6) of the Dark Energy Survey (DES) to analyze their photometric properties. The DES data allow us to study Jupiter Trojans with a fainter magnitude limit than previous studies in a homogeneous survey with $griz$ band measurements. We extract a final catalog of 573 unique Jupiter Trojans. Our sample include 547 asteroids belonging to L5. This is one of the largest analyzed samples for this group. By comparing with the data reported by other surveys we found that the color distribution of L5 Trojans is similar to that of L4 Trojans. We find that L5 Trojans' $g - i$ and $g - r$ colors become less red with fainter absolute magnitudes, a trend also seen in L4 Trojans. Both the L4 and L5 clouds consistently show such a color-size correlation over an absolute magnitude range $11 < H <  18$. We also use DES colors to perform taxonomic classifications. C and P-type asteroids outnumber D-type asteroids in the L5 Trojans DES sample, which have diameters in the 5 - 20 km range. This is consistent with the color-size correlation.  
%The fraction of identified C-type asteroids is also relatively large compared with Trojans at larger sizes. Lastly, we report three potential V- and four potential S-type Jupiter Trojans, but further follow-up is necessary to confirm their taxonomic classes. 
\end{abstract}



\section{Introduction} \label{sec:intro}

The properties of Jupiter Trojans, small bodies that populate the 1:1 mean motion resonance near Jupiter's L4 and L5 Lagrange points, encode important clues about the processes that shaped our solar system and its origins. Recent theories, e.g., the Nice Model \citep{Morbidelli2005}, Grand Tack \citep{Walsh2011}, and Jumping Jupiter \citep{Nesvorny2013,Nesvor2015} support the idea that radial migrations have happened in the early solar system. Under this hypothesis, Jupiter Trojans reached their current orbits by scattering inward from the primordial planetesimal disk as the giant planets migrated outward. Thus, the Trojans may share the same origin as Kuiper belt objects in this scenario. 

The alternative hypothesis suggests that it is also possible for Jupiter Trojans to form in their current locations by capturing planetesimals during the formation of Jupiter \citep{Marzari1998,Fleming2000}. Consequently, their relations with trans-Neptunian objects (TNOs) and other small-body populations, e.g., Hildas and main-belt asteroids (MBAs), contain crucial implications for the solar systems formation hypothesis (TNO: \citealt{Fraser2014,Morbidelli2009}, Hildas: \citealt{Wong2017}, MBAs: \citealt{Yoshida2019}). 

Over the last few decades, numerous observations, experiments, and analyses related to Jupiter Trojans have considerably deepened our understanding of their physical properties, including sizes, colors, and taxonomic types. However, our knowledge of the underlying mechanics and compositions responsible for those properties remains poorly constrained \citep{Wong2019}.
With the upcoming exploration of Lucy spacecraft \citep{Levison_2021}, further analysis of the Jupiter Trojans is an even more compelling task.
%Jupiter Trojans contain information for the initial formation conditions and einformation for environments of solar systems that shaped them afterward. 
%Recently, much progress has been made in understanding the compositions \citep[e.g.,][]{Chang2021,Kalup2021} and evolution \citep[e.g.,][]{Sisto2019}  of Jupiter Trojans, 
%With about 1 million Jupiter Trojans larger than 1 km in diameters, they also represent an important target for understanding connections between the middle solar system and outer solar populations.

Jupiter Trojans have several important features. The color bimodality of Trojans has been claimed in many previous studies in both spectroscopic and photometric surveys \citep{Szab2007, Roig2008, Wong2014, Wong2015}. \citet{Szab2007} analyzed 869 unique Jovian Trojans in the Sloan Digital Sky Survey Moving Object Catalogue third release. They found that the colors of Trojans have small scatters and are correlated with orbital inclination.
\citet{Emery2011} identified two compositional groups in the Jovian Trojan population using near-infrared spectra, which shows two distinct ``red'' and ``less-red'' groups. 
\cite{Wong2014} found that the ``red'' and ``less-red'' groups show different magnitude distributions. \cite{Wong2015} showed that there are more ``less red'' Trojans in $g - i$ colors with decreasing sizes in the L4 clouds. Furthermore, \cite{Wong2019} reported distinct UV spectral shapes between the ``red'' and ``less-red'' Trojans. %Lastly, significant differences exist between L4 and L5 clouds. The L4 cloud contains more large objects with D $>$ 10 km relative to the L5 cloud \citep{Grav2011}. A potential explanation is the passing of an ice giant through the L5 cloud, which scattered many of its Trojans \citep{Nesvorny2013}. In addition to the asymmetry in number, the color distribution \citep{Szab2007}, spectral slope \citep{Roig2008}, and shape distribution \citep{McNeill2021} between L4 and L5 Trojans also contain discrepancies. Possible explanations include the differences in number density and collisional environments between the two clouds. Many studies have examined L4 Trojans' photometry or compared L4 Trojans to a smaller sample of L5 Trojans. To date, a deep, wide-area photometric study focusing on L5 Trojans has been lacking.  

Large-scale photometric surveys can statistically study the color and taxonomic type of Trojans, and these properties are indicators of the surface composition of Jupiter Trojans, or even more, can be used as the parameters to identify possible collisional families \citep{Holt2021}.
Previously, the Sloan Digital Sky Survey (SDSS; \citealt{Ivezi2001, Ivezi2002}), WISE \citep{Mainze2011}, {\tt\string SkyMapper} \citep{Sergeyev2022}, and VISTA (\citealt{Popescu2016}, in near-infrared) have studied surface properties for a large number of Jupiter Trojans. However, these surveys are generally biased toward Trojans with large sizes (usually bigger than diameters of 20 km). Suprime-Cam (SC) and Hyper Suprime-Cam (HSC), which are mounted on the 8-meter class Subaru telescope, have reached deeper magnitudes and have given insights into the magnitude distributions \citep{Uehata2022, Yoshida2017, Yoshida2008, Yoshida2005} and color-magnitude relation of small Trojans \citep[]{Wong2015}. However, those surveys generally lack the multiple band measurements that enable the measurement of taxonomic types. 

This study, carried out with data from the Dark Energy Survey (DES) \citep{DES2005} reaches a deeper magnitude limit, $\sim$ 15 in absolute magnitude $H_{V}$ and correspondingly to $m_{V}\sim$ 22, than the 4th release of Sloan Digital Sky Survey Moving Object Catalogue (SDSS MOC-4; \citealt{Ivezi2001, Ivezi2002}). Also, DES photometry in the $g$, $r$, $i$, and $z$ bands allows the classification of Trojans into different taxonomic types (e.g. \citealt{Carvano2010,Demeo2013}). 
%SDSS MOC-4 is complete up to Trojans with diameters $>$ 20km, and the absolute magnitude limit is $\sim15$, corresponding to a diameter of 4.90 km, assuming a constant albedo of 0.075. 
The goal of this work is to extend our understanding of the Jupiter Trojans' physical properties at the diameters of 5 - 20 km (assuming a constant geometric albedo of 0.07~\citealt{Grav2011}) and use them to shed light on the formation and evolution of Trojans.

%This study will investigate the color and taxonomy of Trojans  the diameters  5 - 20 km for the first time. 

%Thus, a detailed analysis of Jupiter Trojans physical properties is essential to test different models of planet formation and understand the origins and evolution of our solar system. 

%The color dichotomy remains mysterious in the underlying reasons and will be examined in this study \citep{Wong2019}. We shall also demonstrate the color-size correlation in the L5 cloud. Moreover, many studies have examined L4 Trojans' photometry or compared L4 Trojans to a smaller sample of L5 Trojans. To date, a deep, wide-area photometric study focusing on L5 Trojans has been lacking. 

This paper presents the photometry of Jupiter Trojan asteroids observed/imaged in the full six-year data set from DES. The number of identified L5 Jupiter Trojans is substantially higher than L4 Jupiter Trojans. We present the color distributions of Jupiter Trojans for L5 Trojans. We demonstrate the trend of L5 Trojans' absolute magnitudes with colors, and compared the color-size correlation in different surveys. Finally, we classify these Trojans into different taxonomic classes and further discuss the results' implications.  

\section{DES dataset} \label{sec:dataset}

The Dark Energy Survey (DES) \citep{DES2005} was an optical survey carried out between 2013 and 2019 using the Dark Energy Camera (DECam, \citealt{DECam2015}) on the 4-meter Blanco telescope at Cerro Tololo Inter-American Observatory in Chile. The DES consisted of two interleaved surveys: the wide survey, which imaged a 5000 sq.\ deg.\ area centered upon the north galactic cap in the $grizY$ bands to a single-exposure depth of $r\sim 23.8$, and the supernova survey \citep{Bernstein2012}, which imaged ten 3 sq.\ deg.\ DECam fields at approximately weekly intervals in the $griz$ bands. The photometry system is similar but not identical to the SDSS system\footnote{The DECam filter throughput is publicly available at https://noirlab.edu/science/programs/ctio/filters/Dark-Energy-Camera}, therefore we refer the detail of the photometry dataset of DES to \citet{Drlica-Wagner2018D}. 
Though intended primarily as a cosmological survey, the DES's combination of a large survey area, multi-year time baseline, and single-exposure depth make it an outstanding tool for studying our solar system. DES has yielded discoveries of hundreds of new Kuiper Belt Objects \citep{Bernardinelli2022, Bernardinelli2020, Khain2020}, a dwarf planet candidate at 92 AU \citep{DeeDee}, several Neptune Trojans including the first ultra-red member of this population \citep{Lin2019, Gerdes2016}, and a giant Oort cloud comet \citep{CometBB}. Despite the survey's success in discovering new outer Solar System objects, a search for new Jupiter Trojans is prohibitively expensive due to its computational complexity \citep{Bernardinelli2022}. Still, we were able to identify individual detections of known Trojans in the DES data (most Trojans have multiple individual detections). %xNote that DES data also give dynamical information about Trojans; however, it is not used for the investigation of the photometry of Trojans.

The present work makes use of the 107,631 calibrated single-exposure images and catalogs that comprise the full DES six-year (Y6) data set. These images underlie the coadd images and catalogs that were publicly released in January 2021 as DES DR2 \citep{Abbott_2021}.  All $griz$ exposures are 90 s in duration. The ten supernova fields are distributed within the footprint of DES and have longer exposures. 
%\textit{Insert some details about this data set, e.g. number of exposures and number of SE objects.}

\section{Trojans in the DES data} \label{sec:JTmethod}

This section describes how we extract known Jupiter Trojans listed in the Minor Planet Center (MPC) from the DES Y6 data and further clean the data to get a sample of Trojans with reliable photometry. As of 2021 September, the MPC lists 10,437 objects classified as Jupiter Trojans. First, we identify the exposure/CCD combinations in the DES Y6 data that contain known Trojans. Positional uncertainties of identified objects were estimated and required to be smaller than $2\arcsec$ in RA and DEC to ensure accurate matching in our images. Then, we obtain photometry of these identified objects from a catalog of sources detected in individual exposures in the DES Y6 data, excluding objects with nearby stationary objects. We derive absolute magnitudes $H$ of these Trojans. Finally, we further constrain these objects in the number of single exposure and magnitude uncertainties to get a more reliable catalog of Trojans and photometry. Also, some objects listed as known Trojans are not long-term stable, and they are removed as described in Section~\ref{subsec:contrain} below.

%\noindent{\tt\string\begin\{figure*\}} ... {\tt\string\end\{figure*\}}, \\
%\noindent{\tt\string\begin\{table*\}} ... {\tt\string\end\{table*\}}, and \\
%\noindent{\tt\string\begin\{deluxetable*\}} ... %{\tt\string\end\{deluxetable*\}}. \\

\subsection{Identifying Trojans in Y6 of data from DES} \label{subsec:identify}
\begin{figure*}[ht!]
    \includegraphics[width=\linewidth]{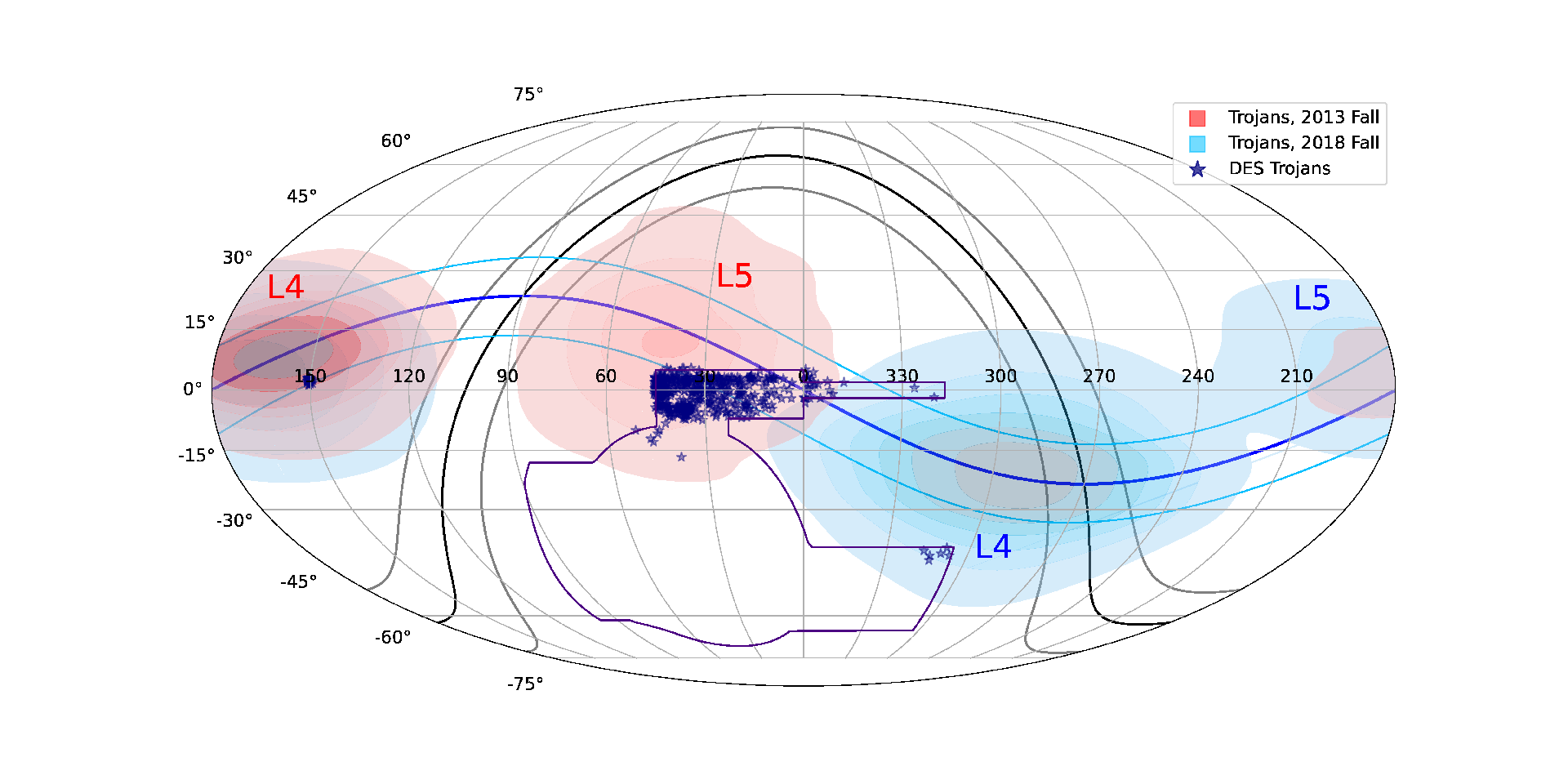}
    \caption{The sky positions of DES observed Jovian Trojans and the DES survey footprint, as well as the ecliptic (blue) and galactic (black) plane. Most of the Trojans belong to L5 camp and were observed during 2013 to 2014. Some L4 Trojans have been picked up in COSMOS field (RA$ \sim 150^{\circ}$) in 2013 or by the main survey in 2018. \label{fig:des_trojan}}
\end{figure*}

The Trojans in the MPC are complete to around absolute magnitude $H$ of 14 on 2019 September 29 \citep{Hendler2020}. Only DES Y6 exposures with ecliptic latitude from -30$^{\circ}$ to 30$^{\circ}$ are searched. We obtain the orbital elements of known Trojans from the MPC and use the \textsc{spacerocks} package \citep{Napier_spacerock} to propagate each Trojan to the epoch of each DES exposure. The \textsc{spacerocks} ephemerides generally agree with JPL Horizons to better than 0.2$\arcsec$ for the numbered minor planets\footnote{See \url{https://github.com/kjnapier/spacerocks/blob/master/notebooks/mpchecker.ipynb}}.
If the ephemeris position falls within the DECam field of view, the object is identified as a potential Trojan in the DES Y6 data. After this search, we obtain 13,732 exposures in DES Y6 data containing 1084 unique Trojans. Their sky positions are shown in Figure~\ref{fig:des_trojan}.

\subsection{Photometry of Trojans}  \label{subsec:photometry}
We cross-match these 1084 objects to sources detected in single exposures in the DES Y6 data with a separation smaller than $2\arcsec$. To avoid contamination from stationary sources that happen to coincide with the predicted position of a Trojan, we use the DES coadd catalog \citep{Abbott_2021} as a veto. This catalog is obtained by co-adding single images in each band acquired throughout the duration of the survey; co-added sources are therefore presumed to be stationary. 
Single-exposure detections that fall within $1\arcsec$ of a coadd catalog source are excluded, a cut that reliably retains moving objects uncontaminated by stationary sources \citep{Bernardinelli2020}. We make an exception for cases where the coadded source was detected in exactly one of its constituent images, as this could arise from a sufficiently bright moving object. At this stage, we reach 888 unique Trojans with 12,057 exposures.

Absolute magnitudes ($H$) of Trojans in each exposure are derived from apparent magnitude using the distances and phase angles at the epoch of exposure. The relation between apparent and absolute magnitude is:

\begin{equation}
m = H + 5~log_{10} (\frac{r \Delta}{d_0^2}) - 2.5~log_{10}~q(\alpha),
\label{eq:a_to_h}
\end{equation}
where $r$ and $\Delta$ are heliocentric and geocentric distance, respectively. The $d_0$ is 1 au, and $q(\alpha)$ is the phase integral. We chose the standard HG model with $G = 0.15$.  

%\textsc{spacerocks} with orbital parameters from MPC, where we estimate orbital information at the epoch of each exposure using \textsc{spacerocks} \citep{Napier_spacerock}. 
Once we derived the $H$ for each object in each epoch in each band, the weighted means of $H$ in each specific band are taken as the final value for absolute magnitude, which is: 

\begin{equation}
H = \frac{\sum_{i=1}^{n} \frac{H_{i}}{\sigma_{i}^{2}} }{\sum_{i=1}^{n} \frac{1}{\sigma_{i}^{2}}},
\label{eq:apmag}
\end{equation} 
where $i$ indicates the individual detection in the specific band for the specific object. We sum over the number of detection of each Trojan in each epoch and each band and the uncertainty is:

\begin{equation}
\sigma  = \sqrt{{\sigma_{0}}^{2} + {(-\frac{2.5}{\ln(10)}\frac{\sigma_{flux}}{flux})}^{2}},
\label{eq:mager}
\end{equation} 
where the $\sigma_{0}$ is the value of zero point magnitude uncertainties and it is usually around 0.002 mag. Through standard error propagation, Equation \ref{eq:mager} includes both zero point magnitude uncertainties and apparent magnitude uncertainties which are estimated using $flux$ and $flux$ uncertainties ($\sigma_{flux}$) from DES Y6 data.  The uncertainty for the absolute magnitudes of each Trojan in each band is:

\begin{equation}
 \overline{\sigma}^{2} =\frac{1 }{\sum_{i=1}^{n} \frac{1}{\sigma_{i}^{2}}}.
\label{eq:mager2}
\end{equation} 
Note that we do not include uncertainty from rotational light-curve variability here since we do not have enough data on each Trojan to estimate it. We discuss of the effect of Trojan's rotations further in section \ref{subsec:rotations}. 

%Absolute magnitudes for each band and each exposure were derived using \textsc{spacerocks} with orbital parameters from MPC, where we estimate orbital information at the epoch of each exposure using \textsc{spacerocks} \citep{Napier_spacerock}. The absolute magnitudes of each Trojan were the weighted average of the absolute magnitudes defined similarly to that of the apparent magnitudes. Its error is taken to be the same as that of the apparent magnitude. 

%Known Jupiter Trojans have well known orbits, so orbital parameters from MPC are reliable.

\subsection{Further constraints on the selected Trojans \label{subsec:contrain}}  

We put constraints on the positional uncertainties to ensure that the selected objects are bona fide Trojans.
The positional uncertainties are estimated using the JPL Horizons system. Uncertainties for every identified object at the time of its exposure time are estimated. The JPL Horizons system gives $3\sigma$ uncertainties around the nominal position in arcseconds. We constrain all Trojans to have positional uncertainties smaller than $2\arcsec$ in both RA and DEC. After this process, we arrive at 9864 individual detections of 775 unique Trojans.

%We arrived at 9864 number of single exposures and 775 unique Jupiter Trojans with zero point magnitude after omitting u and Y bands.

To further improve the quality of our photometry, we require that the number of detections of each Trojan is larger than 1, and the magnitude uncertainty is smaller than 0.1. %Finally, Trojans in MPC are not guaranteed to be stable.
Moreover, the MPC identifies Trojans automatically using their orbital elements, which is not 100 percent reliable\footnote{https://www.minorplanetcenter.net/iau/lists/Trojans.html}. We integrate all 584 objects for 10 million years to make sure that they have stable Trojan orbits. We find eleven objects that are not permanently in resonance, indicating a $\sim$ 2$\%$ contamination rate. We remove these eleven objects from the following analysis.

\subsection{Final catalog of Trojans \label{subsec:finalcata}}  
Finally, we arrive at a final catalog of Jupiter Trojans in the DES Y6 data, which contains 573 unique Trojans. In table~\ref{tab:final}, we present this catalog in a machine-readable format.
%we present the final catalog of Jupiter Trojans in the six-year DES dataset which includes the following columns: name, $H_b$, $\sigma_b$ (the absolute magnitude and corresponding uncertainty in b band), where b belongs to $g$, $r$, $i$, and $z$ and the assigned cloud (L4 or L5). The full version of the final catalog is provided in a machine-readable format. 
Table \ref{tab:number} shows the number of detected Trojans in different bands and the combination of $\{g,i\}$, $\{g,r\}$, $\{r,i\}$, $\{i,z\}$ and, $\{g, r, i\}$ bands and cloud (L4 and L5). This corresponds to 178 L5 and 8 L4 unique Trojans with measurements in all bands. We present the analysis of this dataset in the Section~\ref{sec:result}. 
%L4 Trojans in the DES dataset are less focused due to their paucity in number.

% \begin{deluxetable*}{ccc}
% \tablenum{1}
% \tablecaption{The final catalog of Jupiter Trojans in the six years dataset of DES\label{tab:final}}
% \tablewidth{0pt}
% \tablehead{
% \colhead{Column Name} & \colhead{Unit} & \colhead{Description} 
% }
% \decimalcolnumbers
% \startdata
% Name &  & MPC Designation  \\
% $H_{g}$ & mag & Absolute magnitude in $g$ band  \\
% $\sigma_{g}$ & mag & Uncertainty in $H_{g}$  \\
% $H_{r}$ & mag & Absolute magnitude in $r$ band  \\
% $\sigma_{r}$ & mag & Uncertainty in $H_{r}$  \\
% $H_{i}$ & mag & Absolute magnitude in $i$ band  \\
% $\sigma_{i}$ & mag & Uncertainty in $H_{i}$  \\
% $H_{z}$ & mag & Absolute magnitude  in $z$ band  \\
% $\sigma_{z}$ & mag & Uncertainty in $H_{z}$  \\
% L$_{n}$ &  & Assigned cloud (L4 or L5)  \\
% \enddata
%  \tablecomments{The full version of this table is provided in a machine-readable format. }
% % The Distance is also centered on the decimals.  Note that when using decimal
% % alignment you need to include the {\tt\string\decimals} command before
% % {\tt\string\startdata} and all of the values in that column have to have a
% % space before the next ampersand.}
% \end{deluxetable*}

\begin{deluxetable*}{ccc}
\tablenum{1}
\tablecaption{The final catalog of Jupiter Trojans in the six years dataset of DES\label{tab:final}}
\tablewidth{0pt}
\tablehead{
\colhead{Column Name} & \colhead{Unit} & \colhead{Description} 
}
\decimalcolnumbers
\startdata
Name &  & MPC Designation  \\
$H_{g}$ & mag & Absolute magnitude in $g$ band  \\
$\sigma_{g}$ & mag & Uncertainty in $H_{g}$  \\
$H_{r}$ & mag & Absolute magnitude in $r$ band  \\
$\sigma_{r}$ & mag & Uncertainty in $H_{r}$  \\
$H_{i}$ & mag & Absolute magnitude in $i$ band  \\
$\sigma_{i}$ & mag & Uncertainty in $H_{i}$  \\
$H_{z}$ & mag & Absolute magnitude  in $z$ band  \\
$\sigma_{z}$ & mag & Uncertainty in $H_{z}$  \\
L$_{n}$ &  & Assigned cloud (L4 or L5)  \\
\enddata
 \tablecomments{The full version of this table is provided in a machine-readable format. }
\end{deluxetable*}

\begin{deluxetable}{ccc}
\tablenum{2}
\tablecaption{Number of detected Jupiter Trojans in different bands and cloud\label{tab:number}}
\tablewidth{0pt}
\tablehead{
\colhead{Filter Bands} & \colhead{L5} & \colhead{L4} 
}
%\decimalcolnumbers
\startdata
$g$ & 429 & 14  \\
$r$ & 272 & 14  \\
$i$ & 320 & 21  \\
$z$ & 328 & 18  \\
$g,i$ & 206 & 8  \\
$g,r$ & 206 & 8  \\
$r,i$ & 220 & 10  \\
$i,z$ & 192 & 10  \\
$g,r,i$ & 206 & 8  \\
\enddata
% \tablecomments{This table ``hides'' the third column in the \latex\ when compiled.
% The Distance is also centered on the decimals.  Note that when using decimal
% alignment you need to include the {\tt\string\decimals} command before
% {\tt\string\startdata} and all of the values in that column have to have a
% space before the next ampersand.}
\end{deluxetable}

Some Trojans are not assigned to a cloud by the MPC. We assign these Trojans to L4 (L5) if they lead (trail) Jupiter by more than 20$^{\circ}$ the time of observation.

\section{Results} \label{sec:result}
In this section, we present the results from photometric observations of Trojans in the DES data, including the absolute magnitude distribution, color-color diagram, and correlation between colors and sizes. We further investigate the color-size correlation in a combination of SDSS MOC-4, Subaru, and DES data. To help the comparison among different surveys, the $g$, $r$, $i$, and $z$-band DES photometric magnitudes were converted to the SDSS photometric scheme using the equations in Appendix A.4 of \citet{Drlica-Wagner2018D}. Finally, we present the classification of taxonomic types for Trojans.

%\subsection{Absolute magnitude distribution} \label{subsec:magJT}

%The cumulative distribution of absolute magnitude distribution is shown in Fig.~\ref{fig:magdis}. The absolute magnitude in Johnson V band is converted from DES absolute magnitudes in g and r band using the following equations \citep{Smith2002}:
%\begin{equation}
 %\begin{array}{l}
 % V = g - 0.55\times(g-r)_{SDSS} - 0.03\\
 % (g-r)_{SDSS} = (g-r)_{DES} -0.01\\
 % g_{DES} = g_{SDSS} + 0.001 - 0.075\times(g-r)_{SDSS}
 % \label{eq:coloreq}
 %\end{array}
%\end{equation}
%The cumulative distribution suggests that the magnitude limit is around 15 mag. This is deeper than the 3rd release of the SDSS-MOC, which has a limiting magnitude about 12.3 for known Trojans \citep{Szab2007}.
%\begin{figure}[ht!]
%    \label{fig:magdis}
%    \includegraphics[width=\linewidth]{JT_Color/cumJT.png}
%    \caption{Distribution of absolute magnitude of known Trojans in DES survey in Johnson V band.}
%\end{figure}

\subsection{Color distributions of Jupiter Trojans} \label{subsec:colorJT}
\begin{figure*}[ht!]
    \includegraphics[width=\linewidth]{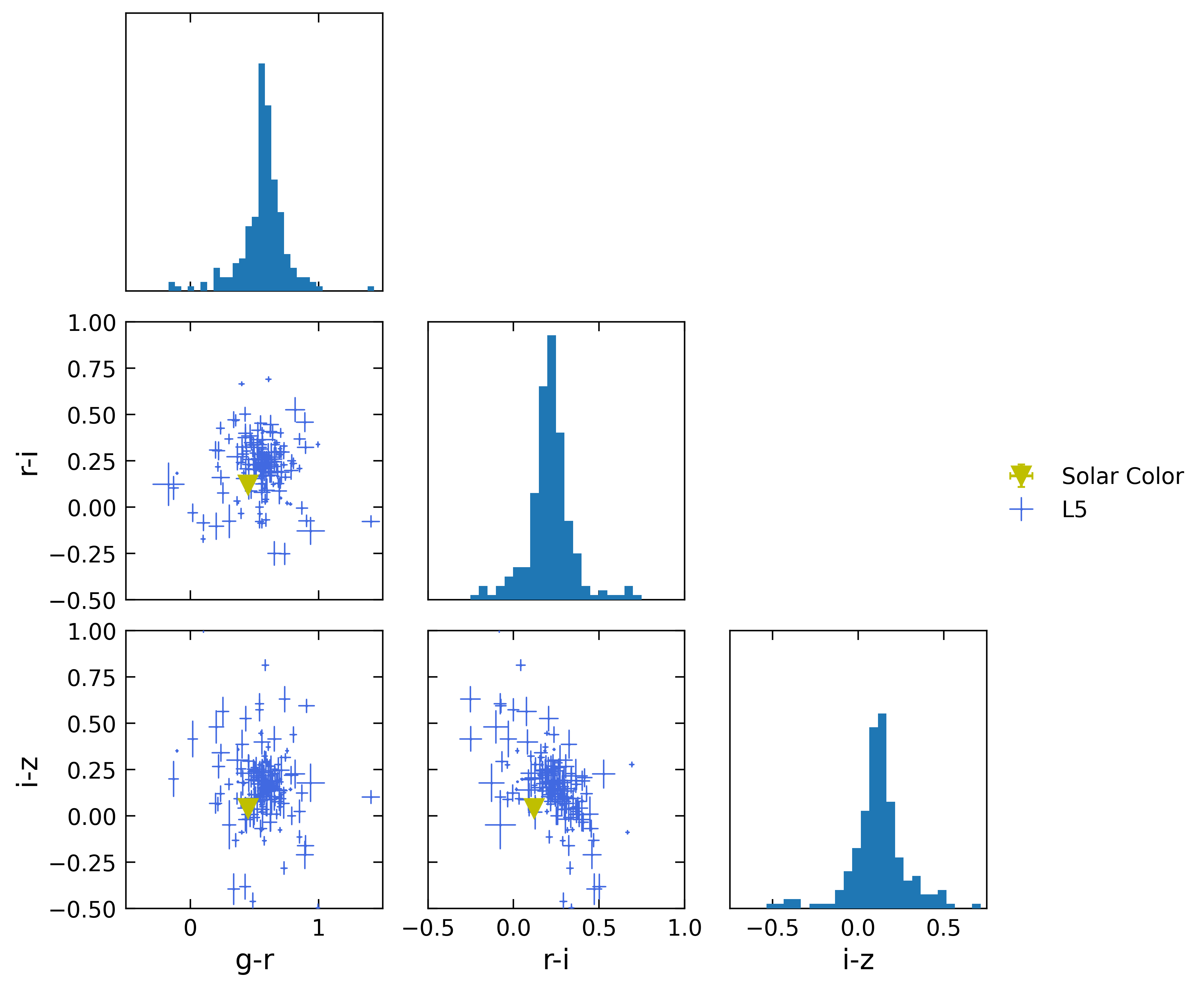}
    \caption{The lower triangle shows Jupiter Trojans' colors for the L5 clouds (blue). The diagonal plots show each color distribution in stacked histograms. The yellow triangle shows the solar color. No bimodality has been found in the colors. L5 Trojans tend to be redder than the solar colors. \label{fig:colordis}}
\end{figure*}

We show the color distribution of Jupiter Trojans in the diagonal of Fig.~\ref{fig:colordis}. The colors in $g-r$, $r-i$, and $i-z$ were restricted to between -2 and 2 to eliminate unphysical red or blue colors. The histogram of colors does not show a bimodality even for bright objects, which has been discernible in the SDSS-MOC 4 data according to~\citet{Wong2015}. We used the unbinned top-hat Kernel Density Estimation method to fit the color distributions and still detect no obvious bimodality. Thus, the lack of bimodality is not an issue with the choice of binning. This could be caused by the smearing of colors due to rotations of Trojans, as was suggested to produce a similar effect in faint L4 Trojans detected by Subaru survey ~\citep{Wong2015}. In particular, the uncertainties introduced by the rotations of Trojans ($\sim$ 0.2 mag) is close to the mean color difference between less-red and red groups, where the latter value is 0.13 mag using Trojans in SDSS-MOC 4~\citep{Wong2015}. The standard deviation for $g-r$, $r-i$, and $i-z$ are 0.18, 0.16, and 0.28 respectively for L5 Trojans. Standard deviations of the color distributions are likely to be enlarged by $\sim 0.1$ mag due to the effects of the rotations of Trojans, which will later be shown in section.~\ref{subsec:rotations}.
%\S \ref{subsec:rotations}). 

Fig. \ref{fig:colordis} lower triangle shows the color-color diagrams of Trojans. The colors are calculated from absolute magnitudes. The color of the sun is overplotted for comparison \citep{solarcolor}. Most L5 Trojans are slightly redder than the sun in both color-magnitude diagrams. The number of L4 Trojans in our sample is too small, below ten in numbers for each color-color plot, to build a reliable sample. Thus, we do not compare L4 and L5 Trojans in this study. The mean $g-r$, $r-i$, and $i-z$ colors of L5 Trojans are 0.56 $\pm$ 0.18, 0.22 $\pm$ 0.16, and 0.17 $\pm$ 0.28 respectively. 

%$The mean colors of L4 Trojans are 0.42 $\pm$ 0.39, 0.25 $\pm$ 0.26, and 0.18 $\pm$ 0.39 for  colors respectively, while those for L5 Trojans are 0.56 $\pm$ 0.18, 0.22 $\pm$ 0.16, and 0.17 $\pm$ 0.28 respectively. The difference in these mean colors between the L4 and L5 clouds does not exceed 0.5 mag. No obvious classifications of red and less-red objects can be found in the color-color diagram. $All colors, including $g-r$, $g-i$, $r-i$, $z-g$, $z-r$, and $i-z$, of the L4 and L5 Trojans were compared using two-sample Kolmogorov-Smirnov (K-S) test. The smallest $p$-value is from the $i-z$ color, which is 0.095. In this paper, we adopt a $p$ value $> 0.01$ for the null hypothesis to be retained. We, thus, cannot reject the possibility that L4 and L5 Trojans' colors are drawn from the same distribution. 
%Though, we cannot say that L4 and L5 Trojans are drawn from different distributions using colors. This conclusion is biased due to the small number of L4 Trojans in our sample.

%We further compare L5 Trojans color distribution with Trojans detected by SDSS and Subaru telescope in section \ref{subsec:compa}.

%The histogram of i-z color have an offset in peak between L4 and L5 Trojans. The median of L5 Trojans i-z colors is 0.12, and that of L4 Trojans is 0.06. This small difference is consistent with i-z colors between L4 and L5 Trojans in ATLAS \citep{McNeill2021}.  

%\subsection{Color color diagrams of Jupiter Trojans} \label{subsec:codiaJT}

%We fitted a linear line  $y = a + b x$ to both color color diagrams, where . In the i-z/r-i diagram, the fitted straight line is $ 0.5917 \pm 0.0435 -1.6531 (r-i) \pm 0.0501$. In the r-i/g-r diagram, the fitted straight line is $732.6399 \pm 0.0795 - 1279.0434(g-r) \pm 0.0403$. The former lines has a negative slopes, while the latter fitted line is close to a vertical line at (g-r) of 0.57. Most Trojans appear to be redder than the colors of the sun. \mynote{need change for colors and fitted lines }

%\begin{figure*}[ht!]
 %   \includegraphics[width=\linewidth]{colorcolor (3).png}
  %  \caption{Color color diagram of Trojans with estimated uncertainties in magnitudes. Left panel is i-z versus r-i, and right panel is r-i versus g-i.  The green triangle is the color of the sun for reference. Black lines in both panels are best-fitted linear line.    \label{fig:colorcolor}}
%end{figure*}

\subsection{Sizes and colors relation}
\label{subsec:sizeJT}
\begin{figure*}[ht!]
    \begin{center}
    \includegraphics[width=12.5 cm]{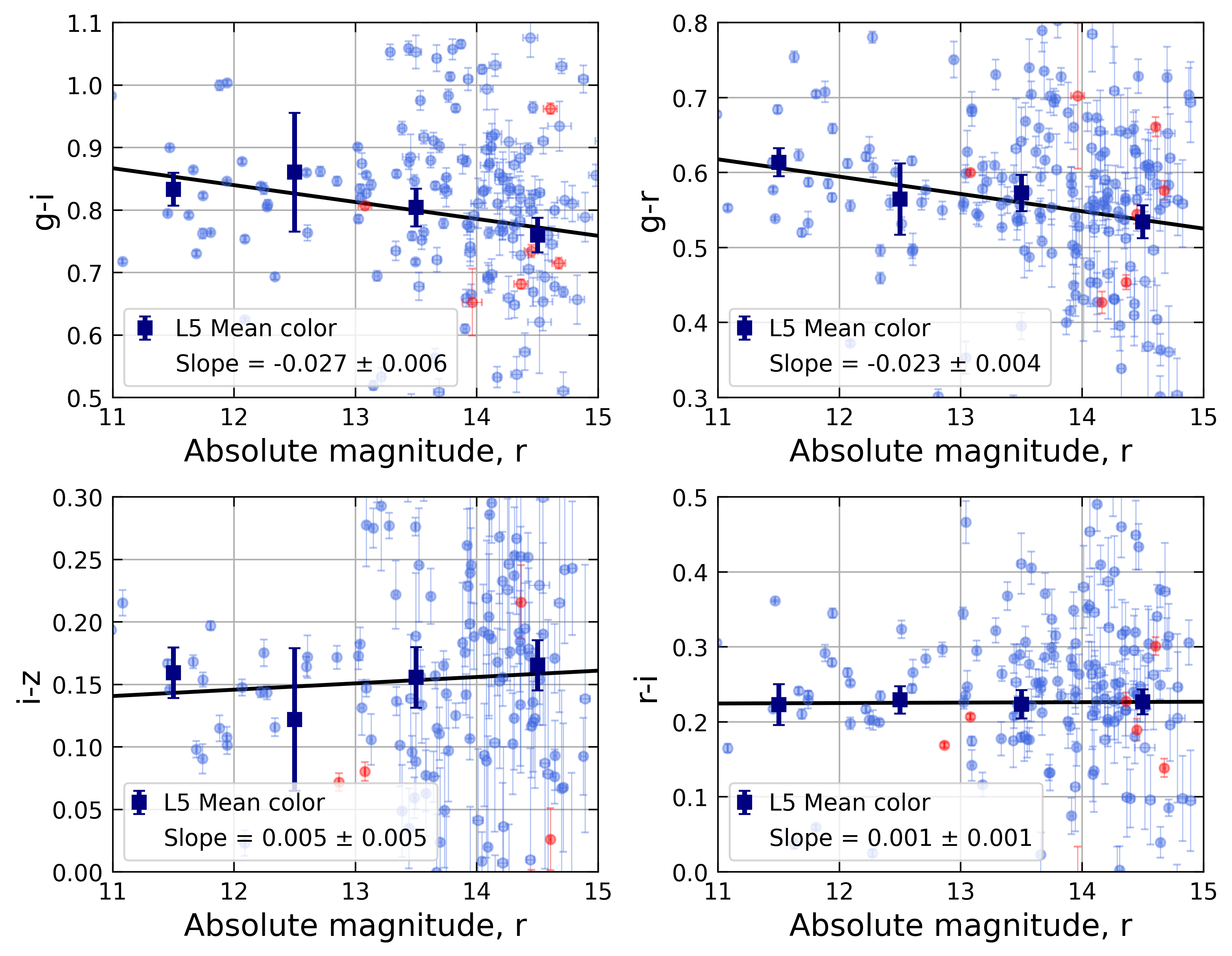}
    \caption{Color magnitude diagram of Trojans (red and blue dot for L5 and L4 Trojans respectively) and mean $g-i$, $g-r$, $i-z$, and $r-i$ colors of L5 Trojans after changing the size of bin to 1 mag (blue squares). The absolute magnitude is used to estimate the size of Trojans. The mean $g-i$ and $g-r$ colors show a trend of becoming less red with decreasing sizes of L5 Trojans. However, the mean $i-z$ and $r-i$ colors do not show such a trend. The black lines are the best-fitted straight lines, excluding outliers.   \label{fig:CMD}}
    \end{center}
\end{figure*}

Here, we use the absolute magnitude to characterize Trojan's size, and study its relation with colors (Fig.~\ref{fig:CMD}). The mean colors of $g-i$, $g-r$, $i-z$, and $r-i$ are the average of each bin of 1 mag, with uncertainties estimated within the same bin. We estimate the uncertainty in mean color as the standard error in each bin. The absolute magnitude of the $r$ band for the mean colors is restricted to 11 - 15, beyond which there are only a few data, so we omit them in this case. We also constrain the colors in $g-i$, $g-r$, $r-i$, and $i-z$ to the ranges of $[-0.5,2]$, $[-0.5,2]$, $[-0.6,1]$, and $[-0.5, 0.8]$ respectively to avoid influences from outliers.
In a linear fit, there is a clear trend of decreasing mean colors in $g-i$ and $g-r$ of L5 Trojans. The best-fitted lines for the $g-i$ and $g-r$ mean colors have slopes of $-0.027 \pm 0.006$ and $-0.023 \pm 0.004$, respectively. In comparison, the slopes for the $i-z$ and $r-i$ mean colors are $0.005 \pm 0.005$ and $0.001 \pm 0.001$ respectively. As a result, the best-fitted lines for $g-i$ and $g-r$ colors are consistent with negative slopes; in contrast, the slopes for $i-z$ and $r-i$ were very close to zero. 
Such a trend does not disappear if the bin size is changed to 0.5 mag. Also, shifting the center of the bin up by 0.5 mag or to the median of the bin does not significantly change the trend. Furthermore, we used F-test to compare the significance of $y = a + bx$ model and $y = a$ model for the data. F-test suggested that there is a clear linear relationship, i.e. $y = a + bx$ model, for $g-i$ and $g-r$ colors with sizes, but not for $i-z$ and $r-i$ colors. The p values for $g-i$, $g-r$, $i-z$, and $r-i$ colors with sizes are 0.099, 0.072, 0.398, and 0.884 respectively. This further confirms our findings of the color-size correlation. Here we adopt a $p$ value to be $<$ 0.1 to reject the null hypothesis of no linear relationship. The correlation breaks down when the magnitude is brighter than 10 because only two objects are in that magnitude range. Such correlations were discovered for the first time in the L5 cloud of Trojans. The $g-i$ colors correlation with sizes agrees with the finding in \cite{Wong2015}, which used only L4 Trojans for their analysis.

%We calculated Spearman's rank correlation between colors and size, and the p-values are bigger than 0.01 for all four colors ($g-i$, $g-r$, $i-z$, and $r-i$) of L5 Trojans. Thus, the correlations between sizes and colors for L5 Trojans are not inconsistent. 

%To further investigate the correlation, we have also fitted lines to the L5 Trojans mean color vs. absolute magnitude. 
%We further investigated the change of sizes with the colors r-i and i-z (middle and lower panel of Fig.~\ref{fig:CMD}). The trends in these two color-magnitude distributions were not more dominant at first glance. To better isolate the trend, we reduced the uncertainties in magnitudes of these two color-magnitudes distributions to 0.05 mag. 
%To further investigate the correlation, we fitted a line of the form $y = a + bx$. The resulted best-fitted lines are shown in Fig.~\ref{fig:CMD}. 

%Outliers with redder than 0.2 i-z colors and bluer than 0.15 r-i colors were excluded in the fitting of linear lines. A linear decreasing in i-z colors can be found at brighter than around 13.5 magnitude. The trend breaks at fainter magnitudes, and this creates a local minimum at around 13 magnitude. The r-i mean colors do not show a decreasing relation with the absolute magnitude, instead the colors were nearly constant with absolute magnitudes. 

However, the DES Trojans were observed at different epochs; the effect of the rotations, which have a typical amplitude of 0.2 mags, may contribute to such a pattern. In Sec.~\ref{subsec:rotations} we use simple simulations to show that the negative slopes are still present even when uncertainties from rotations are included. Also, selection effects may create spurious slopes in $g-i$ and $g-r$ colors with sizes. In particular, a different S/N level in one of two filters could mimic a similar relationship in sizes and colors. For example, if $g$ band has a worse S/N than $r$ or $i$ bands, then redder objects become too faint to be detected in the $g$ band at the faint end of magnitudes. In this scenario, the absence of redder objects at the faint ends of the sample would lead to an artificial trend in $g-i$ and $g-r$ colors and sizes. To address this concern, we examine the S/N of Trojans at all filter bands around the magnitude limits. We found that at around S/N of 10 ($0.08 <\sigma_{m} < 0.14$, where $\sigma_{m}$ is magnitude uncertainty) the magnitude median values are $g=15.2$, $r=14.5$, $i=14.4$, and $z=14.3$. This means that Trojans would have been more reliably detected in the $g$ band than $r$ or $i$ bands, contrary to the potential problematic scenario. Therefore, the color-size correlations are unlikely to be caused by selection effects.

%Further discussion of the effects of rotations on colors and color-size correlation will be in \S \ref{subsec:rotations}. 

%We have also used Gaussian mixture model (GMM) to classify the color magnitude diagram into two sub-populations. As on the first sight, there seem to two groups with different density at brighter and fainter magnitudes for L5 Trojans. The resulted classification of GMM is shown in Figure~\ref{fig:GMM}. It shows that there are two different groups of L5 Trojans in the g-z/g color magnitude diagram. The red group is more concentrated and has more objects with brighter magnitudes. The blue group is more scattered and most of its object is faint. To test the significance of this finding, we have constrained the g-z colors to 0.5 to 1.3. After that, there are still two different groups at brighter and fainter magnitudes. The group at fainter magnitude has a higher density than the group at brighter magnitude. Thus, L5 Trojans tend to be higher in numbers and have more scattered colors at fainter magnitude. This is consistent with the hypothesis that less red Trojans are collisional fragments and there are more less red Trojans sine both red and less red Trojans become less red upon radiation \citep{Wongbrown2016}.

%The scattered colors at fainter magnitudes may be caused by larger uncertainties of magnitudes at fainter magnitudes.   

\subsection{Comparison with Trojans in other surveys}
\label{subsec:compa}
\begin{figure*}[ht!]
    \begin{center}
    \includegraphics[width = 18 cm]{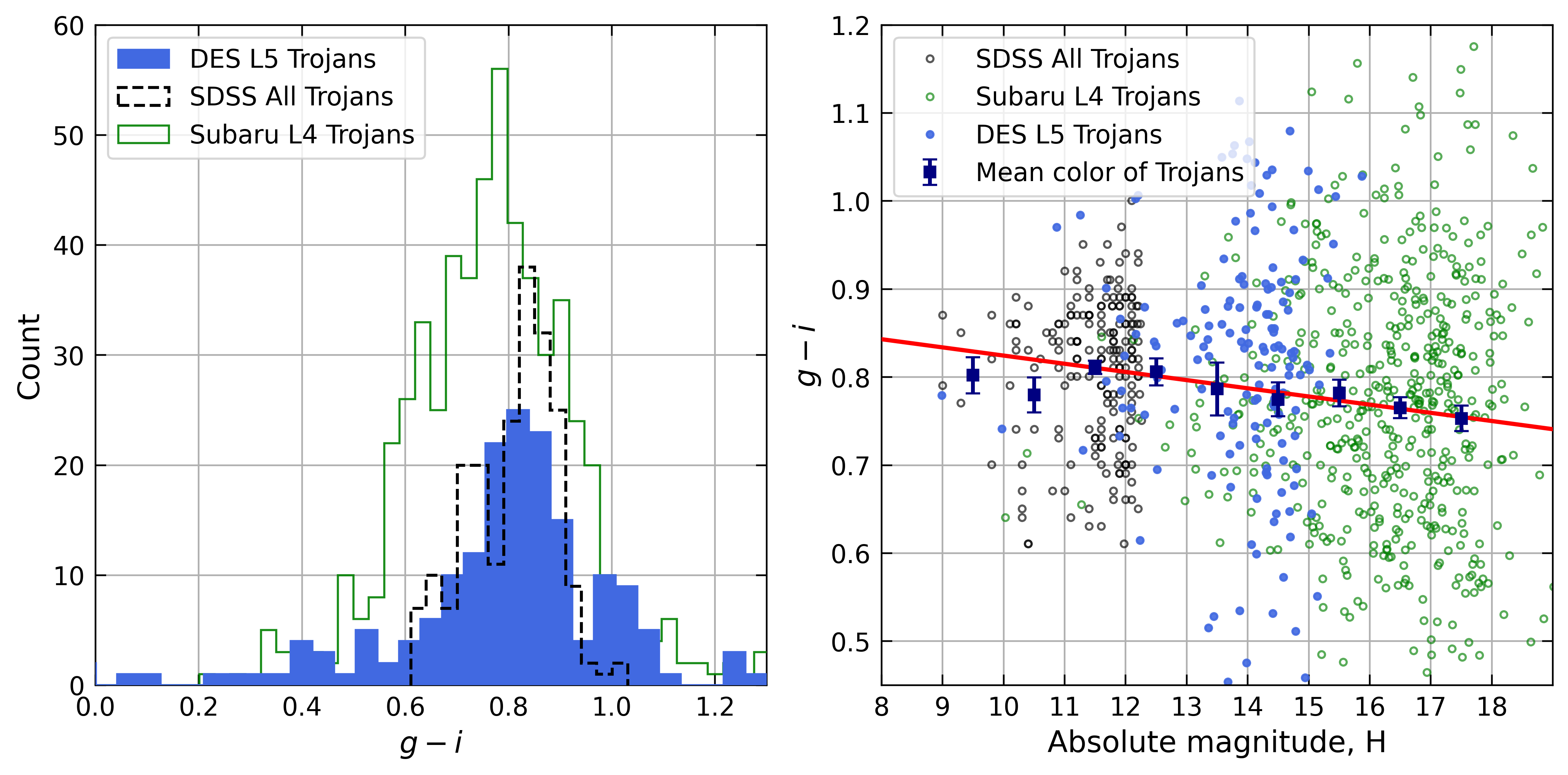}
    \caption{The left panel shows $g-i$ color distributions of Trojans in SDSS-MOC4 catalog, Subaru data, and known Trojans in DES data. DES $g$ and $i$ magnitudes were converted to SDSS magnitudes. The right panel shows the same data in a color absolute magnitude diagram, where the mean of $g-i$ colors in a bin size of 1 mag is overplotted. The fit of mean $g-i$ colors does not include bright Trojans (H $>$ 11). The mean $g-i$ colors show a trend of getting less-red with fainter magnitudes in DES and Subaru data. \label{fig:SDSSDES}}
    \end{center}
\end{figure*}

We compared JTs in this study with Trojans in the SDSS MOC-4 and \cite{Wong2015} (Fig.~\ref{fig:SDSSDES}). We extracted known Trojans in the SDSS MOC-4 catalog. They lie in the regions with distances from 5.04 to 5.4 AU \citep{Demeo2013} and have $e$ $<$ 0.3. The absolute magnitude was constrained to be brighter than 12.3, at which the SDSS MOC-4 catalog is almost complete. SDSS MOC-4 contains more Trojans than DES data due to the larger coverage area. \cite{Wong2015} (hereafter Subaru data) used the Suprime-Cam instrument for the measurement of Trojan colors. The $g-i$ colors in their sample were already calibrated to the SDSS photometric system. We further checked whether the conversion between SDSS and Suprime-Cam magnitudes depends on color, and found that the color-terms are almost negligible (see Appendix~\ref{app:color-term}). Thus, we take $g-i$ colors from \cite{Wong2015} as SDSS $g-i$ colors. At the absolute magnitude interval from 11 to 13, where SDSS and DES data overlap, the mean $g-i$ color difference between SDSS and DES data is around 0.009 $\pm$ 0.04 mag. At absolute magnitude intervals from 13 to 15, the mean $g-i$ color between DES and Subaru data differs by around 0.03 $\pm$ 0.06 mag. The offsets between three different data sets are small compared with the dispersion of the data. However, there could still be contributions from some un-calibrated systematic effects other than color terms between the three photometric systems. Therefore, we conservatively shift the offsets, so that the overlapping absolute magnitude intervals of these three samples have the same mean $g-i$ colors. The following analyses would not be driven by the differences among the three photometric systems.
%To avoid the difference in mean colors contribute to the color-size correlation later, we force the mean color between DES and Subaru data to be the same. 

The left panel of Fig.~\ref{fig:SDSSDES} shows the histograms of L5 Trojans in the DES data, all the Trojans in the SDSS MOC-4 catalog, and L4 Trojans from Subaru data. The mean colors among the three datasets are very close, with a difference smaller than 0.1 mag. The small peak of DES L5 Trojans at $g-i$ around 1 mag is likely an artifact caused by the uncertainties in magnitudes, and it disappears at some other choices of binning. We also note that the DES Trojans colors have a larger scatter than SDSS and Subaru Trojans. We maintain that this is an effect caused by the rotations of Trojans, and it will be discussed in section \ref{subsec:rotations}. Also, KS tests show that $g-i$ distributions in all three data are not compatible with each other.

Fig.~\ref{fig:SDSSDES} right panel shows the mean $g-i$ colors as a function of absolute magnitudes in the $V$ band. The DES reaches a depth in between SDSS and Subaru. We found that the mean $g-i$ colors have a decreasing trend for SDSS, DES, and Subaru data. Mean $g-i$ colors and their uncertainties with a bin size of 1 mag are overplotted. The uncertainty in mean color is estimated in the same way as that of the previous section. Bright Trojans in the SDSS data seem to deviate from this trend ($H$ $>$ 11), as they tend to be bluer than the expected correlation. The agreement of these bright Trojans with the trend is sensitive to variations in bin sizes. It is likely that the strong color-size correlation breaks for these bright objects, consistent with finding in \citet{Szab2007}. Further studies are needed to understand why the correlation breaks for these very big Trojans. Note that in the collisional interpretation, the large objects will not follow such a color-size correlation, since they are expected to not form fragments. Further discussion will be in section~\ref{sec:discussion}. The trend of objects at fainter magnitudes still shows a clear decreasing trend at different bin sizes.
%The bin size was chosen to optimize the visualisation of the mean colors with the trend. Variations of bin sizes show that the overall increasing trend is not affected due to bin sizes.
%This is likely due to the small depth of SDSS data was not enough to cover many smaller objects with bluer $g-i$ colors.
Regardless of the break at brighter magnitudes, the faint end of SDSS Trojans ($H > 11$) is still consistent with the expected color-size correlation. Also, the mean of the Trojan colors in the SDSS data is redder than both DES and Subaru data, with the mean colors in the Subaru data being the less-red. This is also consistent with the correlation. Additionally, we considered SDSS MOC-4 L4 and L5 clouds separately. No significantly different conclusions have been found.

%Additionally, we found that L5 DES Trojans g-i color distribution is consistent with being drawn from the same distribution with SDSS MOC-4 L5 swarm but not L4 swarm on based on two sample K-S test. We found that both L4 and L5 SDSS Trojans g-i color distribution are consistent with L5 DES Trojans after adding simulated uncertainties to SDSS data (further discussion on \S \ref{subsec:rotations}). We maintain that the inconsistency between DES L4 and SDSS L5 Trojans is likely caused by a large scatter of colors in the DES data, resulting from uncorrected light curves amplitudes. However, 

A fitted line for the mean $g-i$ colors, shown as the red line in the right panel of Fig.~\ref{fig:SDSSDES}, has a slope of  $-0.009 \pm 0.001$. Bright Trojans with $H > 11$ are not included in the fit. This slope is around three times smaller than the slope in the $g-i$ color in the DES data ($-0.027 \pm 0.006$). Moreover, we performed a linear fit for L4 Trojans in the Subaru data only with absolute magnitude from 12 to 18 and $g-i$ color from 0.4 to 1.2 to exclude Trojans with color biases and large uncertainties~\citep{Wong2015}. Subaru L4 Trojans have a slope of $-0.010 \pm 0.001$. Similar to the joint data, the slope of Trojans in Subaru data is smaller than that of DES Trojans by around a factor of 3. Nevertheless, a negative slope is still statistically important in the Subaru data within the error bar. Also, both slopes in the red line of Fig.~\ref{fig:SDSSDES} right panel and Subaru data are within three sigma away from the DES derived slope. Here we only consider uncertainties in the colors. Thus, we found a similar relation between colors and sizes in DES and Subaru data. The strong correlation persists from faint Trojans (diameters $\lesssim$ 4 km or $H$ $\gtrsim$ 18) until bright Trojans (diameters $\gtrsim$ 30 km or $H$ $\lesssim$ 11). Even though we used L4 and L5 Trojans data, which have several differences in physical properties, the color-size correlation is present in both clouds with different magnitude ranges. 

  \subsection{Effects of asteroid rotations}
\label{subsec:rotations}
Since the DES color measurements of the Trojans were not simultaneous, the rotation of the objects needs to be taken into account. The DES measured colors can be described as the following equation:

\begin{equation}
C_{obs}  = C_{true} + rot = \bar{C} + \sigma + rot,
\label{eq:obs_color}
\end{equation}
here $C_{obs}$ is the color we measured, $C_{true}$ is the true color of the object, $\bar{C}$ is the mean color of the sample, $\sigma$ is the intrinsic color dispersion from the sample mean of the object, and $rot$ is the rotational effect term, which is the deviation induced by the object rotation. 
The mean colors we calculated in section~\ref{subsec:colorJT} and \ref{subsec:compa} are:
\begin{equation}
\langle C_{obs}\rangle  = \frac{1}{n} \sum_{i=1}^{n}(\bar{C} + \sigma_i + rot_i) = \bar{C} + \langle \sigma \rangle  + \langle rot \rangle.
\end{equation}
Here $n$ is the total number of the sample. By definition, the average of intrinsic color dispersion term $\langle \sigma \rangle$ is zero. If the average of the rotational effect term $\langle rot \rangle$ is also zero, we have $\langle C_{obs}\rangle = \bar{C}$.

It is not possible to distinguish the $\sigma$ and $rot$ from the DES color measurements. Therefore, to test this assumption, we conservatively treated all of the deviations as intrinsic color dispersion and add the additional rotational term to each Trojan. Trojans at the sizes of DES generally have a light curve amplitude of 0.2 mag \citep{Chang2021}. We assumed that the light curves of Trojans follow a sinusoidal curve with amplitudes randomly drawn from a Gaussian distribution at a mean of 0.2 and a standard deviation of 0.1. We randomly sampled a phase of lightcurve from 0 to 2$\pi$ for each observation and add this additional rotation term into the photometric measurements. Applying this step to all the objects, we obtained a new color distribution and a new mean color. After repeating the above steps 100,000 times, we found that the new mean colors agree with the original mean color within $\pm 0.017$. The new standard deviation of the colors tended to increase by $0.079 \pm 0.015$ compared with the original color standard deviation. 

Furthermore, we studied whether the observed decreasing trend of colors with fainter magnitudes is still present with the additional rotation term. We calculated the best-fitted slopes of mean $g-i$ colors vs. absolute magnitudes with a bin size of 1 mag. We found that the slopes tend to stay at a mean of -0.02 with a large standard deviation of 0.02. The negative slope is still present, and the additional rotation did not change the results. A larger standard deviation is expected as we added the extra deviations into the colors. From the above tests, we concluded that the average of rotation term $\langle rot \rangle$ is close to zero, and $\langle C_{obs}\rangle \sim \bar{C}$, which means the mean observed color is very close to the mean color of the Trojan sample. 

The rotational effect also explains the larger color scattering in the DES data. As shown in Fig.~\ref{fig:SDSSDES}, the DES L5 Trojans have $g-i$ colors have some extremes in both very red ($g-r >1$) and very blue ($g-r < 0.6$) ends. In contrast, SDSS MOC-4 Trojans all lie in a very narrow range of colors. A simple explanation is that, unlike the DES data, SDSS colors were taken simultaneously, therefore, the scattering of SDSS colors is pure intrinsic color dispersion ($\sigma$). On the other hand, the scattering of DES colors is intrinsic color dispersion plus a rotational effect, as we described in Equation~\ref{eq:obs_color}. To test this explanation, we ran two samples K-S test between the $g-i$ color dispersion distribution of the SDSS and DES samples and obtained a p-value of 0.004, which means the two $g-i$ color dispersion distributions are likely different from each other. This result was expected, because we compared the $\sigma_{sdss}$ to the $\sigma_{DES} + rot_{DES}$. Then, we added the simulated rotational terms into the SDSS sample. The K-S test returned a p-value of 0.285, which means the color dispersion distributions were now indistinguishable. From the above test, we concluded that the larger color scattering in the DES data was likely induced by the rotational effect. The test also means that any other source of random additional photometric variance with similar uncertainties, e.g., different noise levels on observations of colors, is plausible to account for the difference between SDSS and DES colors.

We note that our function for the amplitudes is not perfect. Some Trojans do have light curve amplitudes larger than the 0.2 mag mean value that we have assumed above. Nevertheless, these larger light curve amplitudes are not common, as only $\sim$5\% of identified Trojans have amplitude larger than 0.4 mag \citep{Mottola2011}. We maintain that our approximation is sufficient for the analysis.

 \subsection{Taxonomic classification of Jupiter Trojans}
\label{subsec:spectral}
Generally, we expect the majority of Trojans to be D-type (red), P-type, or C-type (less-red) asteroids \citep{Demeo2013}. However, most of their classifications are based on Trojans in the SDSS and WISE data, which mostly have diameters $>$ 20 km. The diameters of most Trojans in the DES data are smaller than that limit. This allows us to probe into the spectral types of a large sample of Trojans with diameters from 5 to 20 km.

In this section, we classify the detected Jupiter Trojans into different classes according to their colors following  \citealt{Demeo2013} classification, which used the spectral slope calculated from $g$, $r$, and $i$ reflectance values ($gri$ slope) and $i-z$ colors. The DES photometric magnitudes have been converted to SDSS photometric magnitudes as described in section \ref{subsec:compa}. Only Trojans with measurements in all $griz$ bands were classified, which include 178 L5 Trojans and 8 L4 Trojans as mentioned in section \ref{subsec:finalcata}.

\begin{figure*}[ht!]
    \begin{center}
    \includegraphics[scale=0.8]{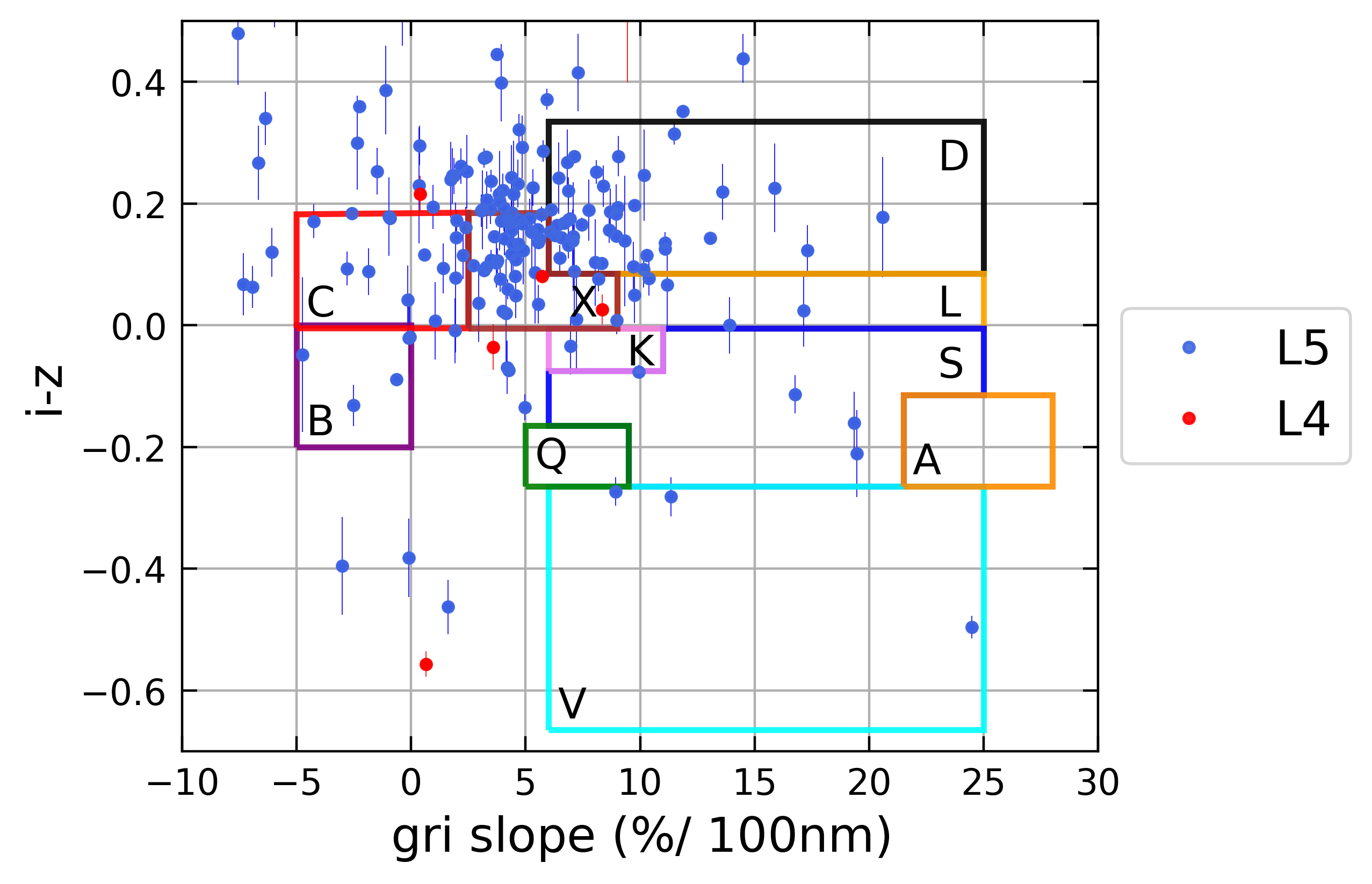}
    \caption{The classification of taxonomy for Jupiter Trojans. Each boundary represents a different class. }
    \label{fig:spectral}
    \end{center}
\end{figure*}

Figure~\ref{fig:spectral} shows the distribution of the 186 Trojans on $gri$ slope vs $i-z$ diagram. The Trojans are mostly located in X and D-type regions, and the X-type region contains three degenerate classes E, M, and P. The large scattering is likely to be caused by the rotational effect (see section~\ref{subsec:rotations}). The center of the distribution is located in the X-type region, which may indicate that in the range of $13 < H < 15$ or in a diameter size range of 5 to 13 km (assuming albedo = 0.07), there are more P-type Trojans than D-type.
Such result is consistent with \citealt{DeMeo2014}, which also shows that there are more P and C-type Trojans than D-type in the smaller size range. Since the P and C-type are less-red than the D-type (shallower $gri$ slope), this result is consistent with what we found: the mean color is less-red for smaller size Trojans. 

The amplitudes of asteroid rotations are generally around 0.2 mag \citep{Mottola2011}. Other than the color uncertainties from asteroid rotation, the intrinsic $i-z$ color uncertainties are usually around 0.04 mag. Figure~\ref{fig:spectral} include the intrinsic $i-z$ color uncertainties. Thus, Trojans with exotic taxonomic types, e.g., S and V-type, should be confirmed with further studies. The difference among C, X, and some D-type Trojans is very subtle, primarily dependent on the $gri$ slope, as seen in Fig.~\ref{fig:spectral}. The uncertainties in slopes average on 2$\%$.

\section{Discussion} \label{sec:discussion}
%In this section, we discuss color-size correlations of DES Trojans and the implication for the formation and evolution of Trojans. 

%\subsection{Color dichotomy and color-size correlation of Trojans}
%\label{subsec:formation}
 Since the fit of L5 Trojans mean $g-i$ colors in Fig. \ref{fig:SDSSDES} is 0.83 mag at $H$ of 9 mag and 0.75 mag at $H$ of 18 mag and agrees very well with the mean $g-i$ colors of the red (mean = 0.86 mag) and less-red (mean = 0.73 mag) populations in SDSS MOC-4 data, which are obtained by fitting a two-peaked Gaussian distribution by \citet{Wong2015}. The increasing fraction of P and C-type asteroids compared with D-type asteroids is also consistent with the increasing number of less-red objects; as P and C-type asteroids have smaller $gri$ slopes than D-type asteroids. 
These two pieces of information hint strongly that two distinct populations with different size distributions and surface properties are responsible for the color bimodality of Trojans, with more P and C-type or less-red asteroids for smaller Trojans. 
%This study thus supports that collision effects are the main causes of these patterns of colours. Consequently, the difference in the shape distribution between L4 and L5 clouds is also supported to be in collision rate has been suggested between L4 and L5 clouds \citep{McNeill2021}. 

%Next, we discuss the color size correlation of Trojans. 
One hypothesis for the color-size correlation is that red populations were converted to the less-red population as they become fragments \citep{Wongbrown2016}, exposing fresher surfaces. They proposed that collisional fragments of both red and less-red groups become less-red in colors due to lack of CH$_{3}$OH and H$_{2}$S. However, further spectroscopic study has not identified any discernible feature in Jupiter Trojans \citep{Wong2019}. %, they are now covered with a thick layer of irradiated mantle after four billion years of space weathering. 
It is also possible that the color dichotomy is mainly caused by the difference in surface properties between the two populations. Since the two populations have different size distributions \citep{Wong2014} and the less-red are more populated than the red group in the smaller end, the color-size correlation reported in \citealt{Wong2015} and this work would be an observational consequence of this fact. 
Nevertheless, whatever mechanisms created the color-size correlation of Jovian Trojans, it must be a general effect between L4 and L5, as both L4 and L5 Trojans share the same color-size trend. 
Finally, the color-size correlation was not obvious in the colors of $i-z$ and $r-i$ in the DES data compared with $g-i$ and $g-r$ colors. Both $i-z$ and $r-i$ colors were almost constant with magnitudes. This is consistent with the fact that the slope of the reflectance spectrum between the red and less-red groups tends to get closer in longer wavelength \citep{Emery2011}.  

%This is also consistent with the increasing P and C-type asteroids. The slope difference in the $g - i$ or $g - r$ colors would be the biggest compared with that in $r-i$ or $i-z$ colors for spectra between D-type asteroids and P and C-type asteroids. 
%Unfortunately, we still have little constraint on the space weathering for different colors and taxonomic classes of asteroids. Thus, we are unable to deduce the conditions of Trojans at the time of their formation from these results. 

\section{Summary} \label{sec:conclusion}

We extracted the known Jovian Trojans from the DES dataset using their orbital parameters in the MPC database. After excluding stationary objects, constraining uncertainties in the positions and photometry, and removing unstable asteroids, we reach a catalog of 547 unique L5 Trojans and 26 unique L4 Trojans. Using this sample, we study the color distributions of known Trojans in DES data with a focus on L5 Trojans. The color of $g-i$ and $g-r$ decreases with smaller sizes of the L5 Trojans, which is similar to the same color-size trend found in the L4 Trojans \citep{Wong2015}. We find no obvious correlations between $r-i$/$i-z$ colors and size of L5 Trojans from the range of $11 < Hr < 15$. Combining the colors derived for DES Trojans with the Trojans from the SDSS MOC-4 data catalog and L4 Trojans from \citet{Wong2015}, we find strong evidence for the color-size correlation of Jovian Trojans, down to absolute magnitudes of $H=18$. Finally, we classify taxonomic types of L5 Trojans and find more potential C and P-type (less-red colored) than D-type (red-colored) asteroids at diameters of 5 - 20 km. The increasing number of C and P-type Trojans is consistent with their color-size correlations, which show that more less-red colored Trojans are at the small-sized end. %Also, identified known Trojans contain a significantly larger fraction of C-type asteroids than the fraction of C-type asteroids of Trojans at larger sizes. 

Future surveys are needed to understand the physical properties and mechanics responsible for the correlations in taxonomic classes/colors and sizes of Jupiter Trojans. We expect that the Lucy mission will greatly enhance our knowledge of the composition of Jupiter Trojans. %(add more descriptions about the slopes )

%Lastly, potential V- and S-type Trojans have been found; however, their identifications require further confirmation.

\section*{Acknowledgments}
This material is based upon work supported by the National Aeronautics and Space Administration under grant No. NNX17AF21G issued through the SSO Planetary Astronomy Program and by the National Science Foundation under grant No. AST-2009096.
We thank Ian Wong and Michael E. Brown for sharing data of Trojans described in \cite{Wong2015}.

Funding for the DES Projects has been provided by the U.S. Department of Energy, the U.S. National Science Foundation, the Ministry of Science and Education of Spain, 
the Science and Technology Facilities Council of the United Kingdom, the Higher Education Funding Council for England, the National Center for Supercomputing 
Applications at the University of Illinois at Urbana-Champaign, the Kavli Institute of Cosmological Physics at the University of Chicago, 
the Center for Cosmology and Astro-Particle Physics at the Ohio State University,
the Mitchell Institute for Fundamental Physics and Astronomy at Texas A\&M University, Financiadora de Estudos e Projetos, 
Funda{\c c}{\~a}o Carlos Chagas Filho de Amparo {\`a} Pesquisa do Estado do Rio de Janeiro, Conselho Nacional de Desenvolvimento Cient{\'i}fico e Tecnol{\'o}gico and 
the Minist{\'e}rio da Ci{\^e}ncia, Tecnologia e Inova{\c c}{\~a}o, the Deutsche Forschungsgemeinschaft and the Collaborating Institutions in the Dark Energy Survey. 

The Collaborating Institutions are Argonne National Laboratory, the University of California at Santa Cruz, the University of Cambridge, Centro de Investigaciones Energ{\'e}ticas, 
Medioambientales y Tecnol{\'o}gicas-Madrid, the University of Chicago, University College London, the DES-Brazil Consortium, the University of Edinburgh, 
the Eidgen{\"o}ssische Technische Hochschule (ETH) Z{\"u}rich, 
Fermi National Accelerator Laboratory, the University of Illinois at Urbana-Champaign, the Institut de Ci{\`e}ncies de l'Espai (IEEC/CSIC), 
the Institut de F{\'i}sica d'Altes Energies, Lawrence Berkeley National Laboratory, the Ludwig-Maximilians Universit{\"a}t M{\"u}nchen and the associated Excellence Cluster Universe, 
the University of Michigan, NSF's NOIRLab, the University of Nottingham, The Ohio State University, the University of Pennsylvania, the University of Portsmouth, 
SLAC National Accelerator Laboratory, Stanford University, the University of Sussex, Texas A\&M University, and the OzDES Membership Consortium.

Based in part on observations at Cerro Tololo Inter-American Observatory at NSF's NOIRLab (NOIRLab Prop. ID 2012B-0001; PI: J. Frieman), which is managed by the Association of Universities for Research in Astronomy (AURA) under a cooperative agreement with the National Science Foundation.

The DES data management system is supported by the National Science Foundation under Grant Numbers AST-1138766 and AST-1536171.
The DES participants from Spanish institutions are partially supported by MICINN under grants ESP2017-89838, PGC2018-094773, PGC2018-102021, SEV-2016-0588, SEV-2016-0597, and MDM-2015-0509, some of which include ERDF funds from the European Union. IFAE is partially funded by the CERCA program of the Generalitat de Catalunya.
Research leading to these results has received funding from the European Research
Council under the European Union's Seventh Framework Program (FP7/2007-2013) including ERC grant agreements 240672, 291329, and 306478.
We  acknowledge support from the Brazilian Instituto Nacional de Ci\^encia
e Tecnologia (INCT) do e-Universo (CNPq grant 465376/2014-2).

This manuscript has been authored by Fermi Research Alliance, LLC under Contract No. DE-AC02-07CH11359 with the U.S. Department of Energy, Office of Science, Office of High Energy Physics.
%\end{acknowledgments}
\vspace{5mm}
\facilities{DECam}

\software{Spacerock \citep{Napier_spacerock}, Astropy \citep{Astropy2013,Astropy2018}, SciPy \citep{2020SciPy-NMeth}, Numpy \citep{numpy}, Matplotlib \citep{matplotlib}, Pandas \citep{pandas}}

\appendix

\section{Color Conversion between Subaru/Suprime-Cam and SDSS Photometric System}
\label{app:color-term}

To convert from Subaru/Suprime-Cam g$_{sc}$, i$_{sc}$ magnitudes to SDSS g$_{sdss}$, i$_{sdss}$, we evaluate the linear color conversions between the two systems using in-frame background stars matched in the SDSS DR12 catalogs. We select the SDSS sources with g$_{sdss} < 21$, i$_{sdss} < 21$, and $0 < (g - i)_{sdss} < 2.5$ to this evaluation. Then, we solve the following equation:
\begin{equation}
m_{sc}  = m_{sdss} + C~(g-i)_{sdss}.
\end{equation}
Here $m_{sc}$ and $m_{sdss}$ are the Subaru and SDSS magnitude, respectively, and C is a linear color-term. 
By solving this equation using the in-frame SDSS sources, which consist with $\sim 10,000$ of individal g-band measurements and $\sim 35,000$ i-band measurements, we find:
\begin{equation}
\label{eq:g}
g_{sc} = g_{sdss} -0.03(g-i)_{sdss},
\end{equation}
and
\begin{equation}
\label{eq:i}
i_{sc} = i_{sdss} -0.02(g-i)_{sdss}.
\end{equation}
Combining Equation~\ref{eq:g} and Equation~\ref{eq:i}, we have
\begin{equation}
(g-i)_{sc} = 0.99(g-i)_{sdss}.
\end{equation}
Since the $g-i < 1$ for most of Trojans, the errors induced by color conversions between Subaru and SDSS photometry systems are less than 1\%. Therefore, we conclude that the effect of color-term is negligible.


\bibliography{maintext}{}

\begin{thebibliography}{}
\expandafter\ifx\csname natexlab\endcsname\relax\def\natexlab#1{#1}\fi
\providecommand{\url}[1]{\href{#1}{#1}}
\providecommand{\dodoi}[1]{doi:~\href{http://doi.org/#1}{\nolinkurl{#1}}}
\providecommand{\doeprint}[1]{\href{http://ascl.net/#1}{\nolinkurl{http://ascl.net/#1}}}
\providecommand{\doarXiv}[1]{\href{https://arxiv.org/abs/#1}{\nolinkurl{https://arxiv.org/abs/#1}}}

\bibitem[{Abbott {et~al.}(2021)Abbott, Adam{\'{o}}w, Aguena, Allam, Amon,
  Annis, Avila, Bacon, Banerji, Bechtol, Becker, Bernstein, Bertin, Bhargava,
  Bridle, Brooks, Burke, Rosell, Kind, Carretero, Castander, Cawthon, Chang,
  Choi, Conselice, Costanzi, Crocce, da~Costa, Davis, Vicente, DeRose, Desai,
  Diehl, Dietrich, Drlica-Wagner, Eckert, Elvin-Poole, Everett, Evrard,
  Ferrero, Fert{\'{e}}, Flaugher, Fosalba, Friedel, Frieman,
  Garc{\'{\i}}a-Bellido, Gaztanaga, Gelman, Gerdes, Giannantonio, Gill, Gruen,
  Gruendl, Gschwend, Gutierrez, Hartley, Hinton, Hollowood, Honscheid, Huterer,
  James, Jeltema, Johnson, Kent, Kron, Kuehn, Kuropatkin, Lahav, Li, Lidman,
  Lin, MacCrann, Maia, Manning, Maloney, March, Marshall, Martini, Melchior,
  Menanteau, Miquel, Morgan, Myles, Neilsen, Ogando, Palmese,
  Paz-Chinch{\'{o}}n, Petravick, Pieres, Plazas, Pond, Rodriguez-Monroy, Romer,
  Roodman, Rykoff, Sako, Sanchez, Santiago, Scarpine, Serrano, Sevilla-Noarbe,
  Smith, Smith, Soares-Santos, Suchyta, Swanson, Tarle, Thomas, To, Tremblay,
  Troxel, Tucker, Turner, Varga, Walker, Wechsler, Weller, Wester, Wilkinson,
  Yanny, Zhang, Nikutta, Fitzpatrick, Jacques, Scott, Olsen, Huang, Herrera,
  Juneau, Nidever, Weaver, Adean, Correia, de~Freitas, Freitas, Singulani, \&
  Vila-Verde}]{Abbott_2021}
Abbott, T. M.~C., Adam{\'{o}}w, M., Aguena, M., {et~al.} 2021, The
  Astrophysical Journal Supplement Series, 255, 20,
  \dodoi{10.3847/1538-4365/ac00b3}

\bibitem[{{Astropy Collaboration} {et~al.}(2013){Astropy Collaboration},
  {Robitaille}, {Tollerud}, {Greenfield}, {Droettboom}, {Bray}, {Aldcroft},
  {Davis}, {Ginsburg}, {Price-Whelan}, {Kerzendorf}, {Conley}, {Crighton},
  {Barbary}, {Muna}, {Ferguson}, {Grollier}, {Parikh}, {Nair}, {Unther},
  {Deil}, {Woillez}, {Conseil}, {Kramer}, {Turner}, {Singer}, {Fox}, {Weaver},
  {Zabalza}, {Edwards}, {Azalee Bostroem}, {Burke}, {Casey}, {Crawford},
  {Dencheva}, {Ely}, {Jenness}, {Labrie}, {Lim}, {Pierfederici}, {Pontzen},
  {Ptak}, {Refsdal}, {Servillat}, \& {Streicher}}]{Astropy2013}
{Astropy Collaboration}, {Robitaille}, T.~P., {Tollerud}, E.~J., {et~al.} 2013,
  \aap, 558, A33, \dodoi{10.1051/0004-6361/201322068}

\bibitem[{{Astropy Collaboration} {et~al.}(2018){Astropy Collaboration},
  {Price-Whelan}, {Sip{\H{o}}cz}, {G{\"u}nther}, {Lim}, {Crawford}, {Conseil},
  {Shupe}, {Craig}, {Dencheva}, {Ginsburg}, {VanderPlas}, {Bradley},
  {P{\'e}rez-Su{\'a}rez}, {de Val-Borro}, {Aldcroft}, {Cruz}, {Robitaille},
  {Tollerud}, {Ardelean}, {Babej}, {Bach}, {Bachetti}, {Bakanov}, {Bamford},
  {Barentsen}, {Barmby}, {Baumbach}, {Berry}, {Biscani}, {Boquien}, {Bostroem},
  {Bouma}, {Brammer}, {Bray}, {Breytenbach}, {Buddelmeijer}, {Burke},
  {Calderone}, {Cano Rodr{\'\i}guez}, {Cara}, {Cardoso}, {Cheedella}, {Copin},
  {Corrales}, {Crichton}, {D'Avella}, {Deil}, {Depagne}, {Dietrich}, {Donath},
  {Droettboom}, {Earl}, {Erben}, {Fabbro}, {Ferreira}, {Finethy}, {Fox},
  {Garrison}, {Gibbons}, {Goldstein}, {Gommers}, {Greco}, {Greenfield},
  {Groener}, {Grollier}, {Hagen}, {Hirst}, {Homeier}, {Horton}, {Hosseinzadeh},
  {Hu}, {Hunkeler}, {Ivezi{\'c}}, {Jain}, {Jenness}, {Kanarek}, {Kendrew},
  {Kern}, {Kerzendorf}, {Khvalko}, {King}, {Kirkby}, {Kulkarni}, {Kumar},
  {Lee}, {Lenz}, {Littlefair}, {Ma}, {Macleod}, {Mastropietro}, {McCully},
  {Montagnac}, {Morris}, {Mueller}, {Mumford}, {Muna}, {Murphy}, {Nelson},
  {Nguyen}, {Ninan}, {N{\"o}the}, {Ogaz}, {Oh}, {Parejko}, {Parley}, {Pascual},
  {Patil}, {Patil}, {Plunkett}, {Prochaska}, {Rastogi}, {Reddy Janga},
  {Sabater}, {Sakurikar}, {Seifert}, {Sherbert}, {Sherwood-Taylor}, {Shih},
  {Sick}, {Silbiger}, {Singanamalla}, {Singer}, {Sladen}, {Sooley},
  {Sornarajah}, {Streicher}, {Teuben}, {Thomas}, {Tremblay}, {Turner},
  {Terr{\'o}n}, {van Kerkwijk}, {de la Vega}, {Watkins}, {Weaver}, {Whitmore},
  {Woillez}, {Zabalza}, \& {Astropy Contributors}}]{Astropy2018}
{Astropy Collaboration}, {Price-Whelan}, A.~M., {Sip{\H{o}}cz}, B.~M., {et~al.}
  2018, \aj, 156, 123, \dodoi{10.3847/1538-3881/aabc4f}

\bibitem[{{Bernardinelli} {et~al.}(2020){Bernardinelli}, {Bernstein}, {Sako},
  {Liu}, {Saunders}, {Khain}, {Lin}, {Gerdes}, {Brout}, {Adams}, {Belyakov},
  {Somasundaram}, {Sharma}, {Locke}, {Franson}, {Becker}, {Napier},
  {Markwardt}, {Annis}, {Abbott}, {Avila}, {Brooks}, {Burke}, {Carnero Rosell},
  {Carrasco Kind}, {Castander}, {da Costa}, {De Vicente}, {Desai}, {Diehl},
  {Doel}, {Everett}, {Flaugher}, {Garc{\'\i}a-Bellido}, {Gruen}, {Gruendl},
  {Gschwend}, {Gutierrez}, {Hollowood}, {James}, {Johnson}, {Johnson},
  {Krause}, {Kuropatkin}, {Maia}, {March}, {Miquel}, {Paz-Chinch{\'o}n},
  {Plazas}, {Romer}, {Rykoff}, {S{\'a}nchez}, {Sanchez}, {Scarpine}, {Serrano},
  {Sevilla-Noarbe}, {Smith}, {Sobreira}, {Suchyta}, {Swanson}, {Tarle},
  {Walker}, {Wester}, {Zhang}, \& {DES Collaboration}}]{Bernardinelli2020}
{Bernardinelli}, P.~H., {Bernstein}, G.~M., {Sako}, M., {et~al.} 2020, \apjs,
  247, 32, \dodoi{10.3847/1538-4365/ab6bd8}

\bibitem[{{Bernardinelli} {et~al.}(2021){Bernardinelli}, {Bernstein}, {Montet},
  {Weryk}, {Wainscoat}, {Aguena}, {Allam}, {Andrade-Oliveira}, {Annis},
  {Avila}, {Bertin}, {Brooks}, {Burke}, {Carnero Rosell}, {Carrasco Kind},
  {Carretero}, {Cawthon}, {Conselice}, {Costanzi}, {da Costa}, {Pereira}, {De
  Vicente}, {Diehl}, {Everett}, {Ferrero}, {Flaugher}, {Frieman},
  {Garc{\'\i}a-Bellido}, {Gaztanaga}, {Gerdes}, {Gruen}, {Gruendl}, {Gschwend},
  {Gutierrez}, {Hinton}, {Hollowood}, {Honscheid}, {James}, {Kuehn},
  {Kuropatkin}, {Lahav}, {Maia}, {Marshall}, {Menanteau}, {Miquel}, {Morgan},
  {Ogando}, {Paz-Chinch{\'o}n}, {Pieres}, {Malag{\'o}n}, {Rodriguez-Monroy},
  {Romer}, {Roodman}, {Sanchez}, {Schubnell}, {Serrano}, {Sevilla-Noarbe},
  {Smith}, {Soares-Santos}, {Suchyta}, {Swanson}, {Tarle}, {To}, {Troxel},
  {Varga}, {Walker}, {Zhang}, \& {DES Collaboration}}]{CometBB}
{Bernardinelli}, P.~H., {Bernstein}, G.~M., {Montet}, B.~T., {et~al.} 2021,
  \apjl, 921, L37, \dodoi{10.3847/2041-8213/ac32d3}

\bibitem[{{Bernardinelli} {et~al.}(2022){Bernardinelli}, {Bernstein}, {Sako},
  {Yanny}, {Aguena}, {Allam}, {Andrade-Oliveira}, {Bertin}, {Brooks},
  {Buckley-Geer}, {Burke}, {Rosell}, {Carrasco Kind}, {Carretero}, {Conselice},
  {Costanzi}, {da Costa}, {De Vicente}, {Desai}, {Diehl}, {Dietrich}, {Doel},
  {Eckert}, {Everett}, {Ferrero}, {Flaugher}, {Fosalba}, {Frieman},
  {Garc{\'\i}a-Bellido}, {Gerdes}, {Gruen}, {Gruendl}, {Gschwend}, {Hinton},
  {Hollowood}, {Honscheid}, {James}, {Kent}, {Kuehn}, {Kuropatkin}, {Lahav},
  {Maia}, {March}, {Menanteau}, {Miquel}, {Morgan}, {Myles}, {Ogando},
  {Palmese}, {Paz-Chinch{\'o}n}, {Pieres}, {Malag{\'o}n}, {Romer}, {Roodman},
  {Sanchez}, {Scarpine}, {Schubnell}, {Serrano}, {Sevilla-Noarbe}, {Smith},
  {Soares-Santos}, {Suchyta}, {Swanson}, {Tarle}, {To}, {Varga}, \&
  {Walker}}]{Bernardinelli2022}
{Bernardinelli}, P.~H., {Bernstein}, G.~M., {Sako}, M., {et~al.} 2022, \apjs,
  258, 41, \dodoi{10.3847/1538-4365/ac3914}

\bibitem[{{Bernstein} {et~al.}(2012){Bernstein}, {Kessler}, {Kuhlmann},
  {Biswas}, {Kovacs}, {Aldering}, {Crane}, {D'Andrea}, {Finley}, {Frieman},
  {Hufford}, {Jarvis}, {Kim}, {Marriner}, {Mukherjee}, {Nichol}, {Nugent},
  {Parkinson}, {Reis}, {Sako}, {Spinka}, \& {Sullivan}}]{Bernstein2012}
{Bernstein}, J.~P., {Kessler}, R., {Kuhlmann}, S., {et~al.} 2012, \apj, 753,
  152, \dodoi{10.1088/0004-637X/753/2/152}

\bibitem[{{Carvano} {et~al.}(2010){Carvano}, {Hasselmann}, {Lazzaro}, \&
  {Moth{\'e}-Diniz}}]{Carvano2010}
{Carvano}, J.~M., {Hasselmann}, P.~H., {Lazzaro}, D., \& {Moth{\'e}-Diniz}, T.
  2010, \aap, 510, A43, \dodoi{10.1051/0004-6361/200913322}

\bibitem[{{Chang} {et~al.}(2021){Chang}, {Chen}, {Fraser}, {Yoshida}, {Lehner},
  {Wang}, {Kavelaars}, {Pike}, {Alexandersen}, {Ito}, {Choi}, {Granados
  Contreras}, {Jeongahn}, {Ji}, {Kim}, {Lawler}, {Li}, {Lin}, {Sofia Lykawka},
  {Moon}, {More}, {Mu{\~n}oz-Guti{\'e}rrez}, {Ohtsuki}, {Terai}, {Urakawa},
  {Zhang}, {Zhao}, {Zhou}, \& {Fossil Collaboration}}]{Chang2021}
{Chang}, C.-K., {Chen}, Y.-T., {Fraser}, W.~C., {et~al.} 2021, \psj, 2, 191,
  \dodoi{10.3847/PSJ/ac13a4}

\bibitem[{{DeMeo} \& {Carry}(2013)}]{Demeo2013}
{DeMeo}, F.~E., \& {Carry}, B. 2013, \icarus, 226, 723,
  \dodoi{10.1016/j.icarus.2013.06.027}

\bibitem[{{DeMeo} \& {Carry}(2014)}]{DeMeo2014}
---. 2014, \nat, 505, 629, \dodoi{10.1038/nature12908}

\bibitem[{{DES Collaboration}(2005)}]{DES2005}
{DES Collaboration}. 2005, ArXiv e-prints.
\newblock \doarXiv{astro-ph/0510346}

\bibitem[{{Drlica-Wagner} {et~al.}(2018){Drlica-Wagner}, {Sevilla-Noarbe},
  {Rykoff}, {Gruendl}, {Yanny}, {Tucker}, {Hoyle}, {Carnero Rosell},
  {Bernstein}, {Bechtol}, {Becker}, {Benoit-L{\'e}vy}, {Bertin}, {Carrasco
  Kind}, {Davis}, {de Vicente}, {Diehl}, {Gruen}, {Hartley}, {Leistedt}, {Li},
  {Marshall}, {Neilsen}, {Rau}, {Sheldon}, {Smith}, {Troxel}, {Wyatt}, {Zhang},
  {Abbott}, {Abdalla}, {Allam}, {Banerji}, {Brooks}, {Buckley-Geer}, {Burke},
  {Capozzi}, {Carretero}, {Cunha}, {D'Andrea}, {da Costa}, {DePoy}, {Desai},
  {Dietrich}, {Doel}, {Evrard}, {Fausti Neto}, {Flaugher}, {Fosalba},
  {Frieman}, {Garc{\'\i}a-Bellido}, {Gerdes}, {Giannantonio}, {Gschwend},
  {Gutierrez}, {Honscheid}, {James}, {Jeltema}, {Kuehn}, {Kuhlmann},
  {Kuropatkin}, {Lahav}, {Lima}, {Lin}, {Maia}, {Martini}, {McMahon},
  {Melchior}, {Menanteau}, {Miquel}, {Nichol}, {Ogando}, {Plazas}, {Romer},
  {Roodman}, {Sanchez}, {Scarpine}, {Schindler}, {Schubnell}, {Smith}, {Smith},
  {Soares-Santos}, {Sobreira}, {Suchyta}, {Tarle}, {Vikram}, {Walker},
  {Wechsler}, {Zuntz}, \& {DES Collaboration}}]{Drlica-Wagner2018D}
{Drlica-Wagner}, A., {Sevilla-Noarbe}, I., {Rykoff}, E.~S., {et~al.} 2018,
  \apjs, 235, 33, \dodoi{10.3847/1538-4365/aab4f5}

\bibitem[{{Emery} {et~al.}(2011){Emery}, {Burr}, \& {Cruikshank}}]{Emery2011}
{Emery}, J.~P., {Burr}, D.~M., \& {Cruikshank}, D.~P. 2011, \aj, 141, 25,
  \dodoi{10.1088/0004-6256/141/1/25}

\bibitem[{Flaugher {et~al.}(2015)Flaugher, Diehl, Honscheid, Abbott, Alvarez,
  Angstadt, Annis, Antonik, Ballester, Beaufore, Bernstein, Bernstein, Bigelow,
  Bonati, Boprie, Brooks, Buckley-Geer, Campa, Cardiel-Sas, Castander,
  Castilla, Cease, Cela-Ruiz, Chappa, Chi, Cooper, da~Costa, Dede, Derylo,
  DePoy, de~Vicente, Doel, Drlica-Wagner, Eiting, Elliott, Emes, Estrada, Neto,
  Finley, Flores, Frieman, Gerdes, Gladders, Gregory, Gutierrez, Hao, Holland,
  Holm, Huffman, Jackson, James, Jonas, Karcher, Karliner, Kent, Kessler,
  Kozlovsky, Kron, Kubik, Kuehn, Kuhlmann, Kuk, Lahav, Lathrop, Lee, Levi,
  Lewis, Li, Mandrichenko, Marshall, Martinez, Merritt, Miquel, Mu{\~n}oz,
  Neilsen, Nichol, Nord, Ogando, Olsen, Palaio, Patton, Peoples, Plazas, Rauch,
  Reil, Rheault, Roe, Rogers, Roodman, Sanchez, Scarpine, Schindler, Schmidt,
  Schmitt, Schubnell, Schultz, Schurter, Scott, Serrano, Shaw, Smith,
  Soares-Santos, Stefanik, Stuermer, Suchyta, Sypniewski, Tarle, Thaler, Tighe,
  Tran, Tucker, Walker, Wang, Watson, Weaverdyck, Wester, Woods, Yanny, \&
  Collaboration}]{DECam2015}
Flaugher, B., Diehl, H.~T., Honscheid, K., {et~al.} 2015, The Astronomical
  Journal, 150, 150.
\newblock \doarXiv{1504.02900}

\bibitem[{{Fleming} \& {Hamilton}(2000)}]{Fleming2000}
{Fleming}, H.~J., \& {Hamilton}, D.~P. 2000, \icarus, 148, 479,
  \dodoi{10.1006/icar.2000.6523}

\bibitem[{{Fraser} {et~al.}(2014){Fraser}, {Brown}, {Morbidelli}, {Parker}, \&
  {Batygin}}]{Fraser2014}
{Fraser}, W.~C., {Brown}, M.~E., {Morbidelli}, A., {Parker}, A., \& {Batygin},
  K. 2014, \apj, 782, 100, \dodoi{10.1088/0004-637X/782/2/100}

\bibitem[{{Gerdes} {et~al.}(2016){Gerdes}, {Jennings}, {Bernstein}, {Sako},
  {Adams}, {Goldstein}, {Kessler}, {Hamilton}, {Abbott}, {Abdalla}, {Allam},
  {Benoit-L{\'e}vy}, {Bertin}, {Brooks}, {Buckley-Geer}, {Burke}, {Capozzi},
  {Carnero Rosell}, {Carrasco Kind}, {Carretero}, {Cunha}, {D'Andrea}, {da
  Costa}, {DePoy}, {Desai}, {Dietrich}, {Doel}, {Eifler}, {Fausti Neto},
  {Flaugher}, {Frieman}, {Gaztanaga}, {Gruen}, {Gruendl}, {Gutierrez},
  {Honscheid}, {James}, {Kuehn}, {Kuropatkin}, {Lahav}, {Li}, {Maia}, {March},
  {Martini}, {Miller}, {Miquel}, {Nichol}, {Nord}, {Ogando}, {Plazas}, {Romer},
  {Roodman}, {Sanchez}, {Santiago}, {Schubnell}, {Sevilla-Noarbe}, {Smith},
  {Soares-Santos}, {Sobreira}, {Suchyta}, {Swanson}, {Tarl{\'e}}, {Thaler},
  {Walker}, {Wester}, {Zhang}, \& {DES Collaboration}}]{Gerdes2016}
{Gerdes}, D.~W., {Jennings}, R.~J., {Bernstein}, G.~M., {et~al.} 2016, \aj,
  151, 39, \dodoi{10.3847/0004-6256/151/2/39}

\bibitem[{{Gerdes} {et~al.}(2017){Gerdes}, {Sako}, {Hamilton}, {Zhang},
  {Khain}, {Becker}, {Annis}, {Wester}, {Bernstein}, {Scheibner}, {Zullo},
  {Adams}, {Bergin}, {Walker}, {Mueller}, {Abbott}, {Abdalla}, {Allam},
  {Bechtol}, {Benoit-L{\'e}vy}, {Bertin}, {Brooks}, {Burke}, {Carnero Rosell},
  {Carrasco Kind}, {Carretero}, {Cunha}, {da Costa}, {Desai}, {Diehl},
  {Eifler}, {Flaugher}, {Frieman}, {Garc{\'\i}a-Bellido}, {Gaztanaga},
  {Goldstein}, {Gruen}, {Gschwend}, {Gutierrez}, {Honscheid}, {James}, {Kent},
  {Krause}, {Kuehn}, {Kuropatkin}, {Lahav}, {Li}, {Maia}, {March}, {Marshall},
  {Martini}, {Menanteau}, {Miquel}, {Nichol}, {Plazas}, {Romer}, {Roodman},
  {Sanchez}, {Sevilla-Noarbe}, {Smith}, {Smith}, {Soares-Santos}, {Sobreira},
  {Suchyta}, {Swanson}, {Tarle}, {Tucker}, {Zhang}, \& {DES
  Collaboration}}]{DeeDee}
{Gerdes}, D.~W., {Sako}, M., {Hamilton}, S., {et~al.} 2017, \apjl, 839, L15,
  \dodoi{10.3847/2041-8213/aa64d8}

\bibitem[{{Grav} {et~al.}(2011){Grav}, {Mainzer}, {Bauer}, {Masiero}, {Spahr},
  {McMillan}, {Walker}, {Cutri}, {Wright}, {Eisenhardt}, {Blauvelt}, {DeBaun},
  {Elsbury}, {Gautier}, {Gomillion}, {Hand}, \& {Wilkins}}]{Grav2011}
{Grav}, T., {Mainzer}, A.~K., {Bauer}, J., {et~al.} 2011, \apj, 742, 40,
  \dodoi{10.1088/0004-637X/742/1/40}

\bibitem[{Harris {et~al.}(2020)Harris, Millman, van~der Walt, Gommers,
  Virtanen, Cournapeau, Wieser, Taylor, Berg, Smith, Kern, Picus, Hoyer, van
  Kerkwijk, Brett, Haldane, del R{\'{i}}o, Wiebe, Peterson,
  G{\'{e}}rard-Marchant, Sheppard, Reddy, Weckesser, Abbasi, Gohlke, \&
  Oliphant}]{numpy}
Harris, C.~R., Millman, K.~J., van~der Walt, S.~J., {et~al.} 2020, Nature, 585,
  357, \dodoi{10.1038/s41586-020-2649-2}

\bibitem[{{Hendler} \& {Malhotra}(2020)}]{Hendler2020}
{Hendler}, N.~P., \& {Malhotra}, R. 2020, \psj, 1, 75,
  \dodoi{10.3847/PSJ/abbe25}

\bibitem[{Holmberg {et~al.}(2006)Holmberg, Flynn, \& Portinari}]{solarcolor}
Holmberg, J., Flynn, C., \& Portinari, L. 2006, Monthly Notices of the Royal
  Astronomical Society, 367, 449, \dodoi{10.1111/j.1365-2966.2005.09832.x}

\bibitem[{{Holt} {et~al.}(2021){Holt}, {Horner}, {Nesvorn{\'y}}, {King},
  {Popescu}, {Carter}, \& {Tylor}}]{Holt2021}
{Holt}, T.~R., {Horner}, J., {Nesvorn{\'y}}, D., {et~al.} 2021, \mnras, 504,
  1571, \dodoi{10.1093/mnras/stab894}

\bibitem[{Hunter(2007)}]{matplotlib}
Hunter, J.~D. 2007, Computing in Science \& Engineering, 9, 90,
  \dodoi{10.1109/MCSE.2007.55}

\bibitem[{{Ivezi{\'c}} {et~al.}(2001){Ivezi{\'c}}, {Tabachnik}, {Rafikov},
  {Lupton}, {Quinn}, {Hammergren}, {Eyer}, {Chu}, {Armstrong}, {Fan},
  {Finlator}, {Geballe}, {Gunn}, {Hennessy}, {Knapp}, {Leggett}, {Munn},
  {Pier}, {Rockosi}, {Schneider}, {Strauss}, {Yanny}, {Brinkmann}, {Csabai},
  {Hindsley}, {Kent}, {Lamb}, {Margon}, {McKay}, {Smith}, {Waddel}, {York}, \&
  {SDSS Collaboration}}]{Ivezi2001}
{Ivezi{\'c}}, {\v{Z}}., {Tabachnik}, S., {Rafikov}, R., {et~al.} 2001, \aj,
  122, 2749, \dodoi{10.1086/323452}

\bibitem[{{Ivezi{\'c}} {et~al.}(2002){Ivezi{\'c}}, {Lupton}, {Juri{\'c}},
  {Tabachnik}, {Quinn}, {Gunn}, {Knapp}, {Rockosi}, \& {Brinkmann}}]{Ivezi2002}
{Ivezi{\'c}}, {\v{Z}}., {Lupton}, R.~H., {Juri{\'c}}, M., {et~al.} 2002, \aj,
  124, 2943, \dodoi{10.1086/344077}

\bibitem[{{Khain} {et~al.}(2020){Khain}, {Becker}, {Lin}, {Gerdes}, {Adams},
  {Bernardinelli}, {Bernstein}, {Franson}, {Markwardt}, {Hamilton}, {Napier},
  {Sako}, {Abbott}, {Avila}, {Bertin}, {Brooks}, {Buckley-Geer}, {Burke},
  {Carnero Rosell}, {Carrasco Kind}, {Carretero}, {Costa}, {Vicente}, {Desai},
  {Diehl}, {Doel}, {Flaugher}, {Frieman}, {Garc{\'\i}a-Bellido}, {Gaztanaga},
  {Gruen}, {Gruendl}, {Gschwend}, {Gutierrez}, {Hollowood}, {Honscheid},
  {James}, {Kuropatkin}, {Maia}, {Marshall}, {Menanteau}, {Miller}, {Miquel},
  {Plazas}, {Sanchez}, {Scarpine}, {Schubnell}, {Sevilla-Noarbe}, {Smith},
  {Sobreira}, {Suchyta}, {Swanson}, {Tarle}, {Walker}, {Wester}, \& {Dark
  Energy Survey Collaboration}}]{Khain2020}
{Khain}, T., {Becker}, J.~C., {Lin}, H.~W., {et~al.} 2020, \aj, 159, 133,
  \dodoi{10.3847/1538-3881/ab7002}

\bibitem[{Levison {et~al.}(2021)Levison, Olkin, Noll, Marchi, III, Bierhaus,
  Binzel, Bottke, Britt, Brown, Buie, Christensen, Emery, Grundy, Hamilton,
  Howett, Mottola, Pätzold, Reuter, Spencer, Statler, Stern, Sunshine, Weaver,
  \& Wong}]{Levison_2021}
Levison, H.~F., Olkin, C.~B., Noll, K.~S., {et~al.} 2021, The Planetary Science
  Journal, 2, 171, \dodoi{10.3847/psj/abf840}

\bibitem[{{Lin} {et~al.}(2019){Lin}, {Gerdes}, {Hamilton}, {Adams},
  {Bernstein}, {Sako}, {Bernadinelli}, {Tucker}, {Allam}, {Becker}, {Khain},
  {Markwardt}, {Franson}, {Abbott}, {Annis}, {Avila}, {Brooks}, {Carnero
  Rosell}, {Carrasco Kind}, {Cunha}, {D'Andrea}, {da Costa}, {De Vicente},
  {Doel}, {Eifler}, {Flaugher}, {Garc{\'\i}a-Bellido}, {Hollowood},
  {Honscheid}, {James}, {Kuehn}, {Kuropatkin}, {Maia}, {Marshall}, {Miquel},
  {Plazas}, {Romer}, {Sanchez}, {Scarpine}, {Sevilla-Noarbe}, {Smith}, {Smith},
  {Soares-Santos}, {Sobreira}, {Suchyta}, {Tarle}, {Walker}, \&
  {Wester}}]{Lin2019}
{Lin}, H.~W., {Gerdes}, D.~W., {Hamilton}, S.~J., {et~al.} 2019, \icarus, 321,
  426, \dodoi{10.1016/j.icarus.2018.12.006}

\bibitem[{{Mainzer} {et~al.}(2011){Mainzer}, {Bauer}, {Grav}, {Masiero},
  {Cutri}, {Dailey}, {Eisenhardt}, {McMillan}, {Wright}, {Walker}, {Jedicke},
  {Spahr}, {Tholen}, {Alles}, {Beck}, {Brandenburg}, {Conrow}, {Evans},
  {Fowler}, {Jarrett}, {Marsh}, {Masci}, {McCallon}, {Wheelock}, {Wittman},
  {Wyatt}, {DeBaun}, {Elliott}, {Elsbury}, {Gautier}, {Gomillion}, {Leisawitz},
  {Maleszewski}, {Micheli}, \& {Wilkins}}]{Mainze2011}
{Mainzer}, A., {Bauer}, J., {Grav}, T., {et~al.} 2011, \apj, 731, 53,
  \dodoi{10.1088/0004-637X/731/1/53}

\bibitem[{{Marzari} \& {Scholl}(1998)}]{Marzari1998}
{Marzari}, F., \& {Scholl}, H. 1998, \icarus, 131, 41,
  \dodoi{10.1006/icar.1997.5841}

\bibitem[{McKinney {et~al.}(2010)}]{pandas}
McKinney, W., {et~al.} 2010, in Proceedings of the 9th Python in Science
  Conference, Vol. 445, Austin, TX, 51--56

\bibitem[{{Morbidelli} {et~al.}(2009){Morbidelli}, {Levison}, {Bottke},
  {Dones}, \& {Nesvorn{\'y}}}]{Morbidelli2009}
{Morbidelli}, A., {Levison}, H.~F., {Bottke}, W.~F., {Dones}, L., \&
  {Nesvorn{\'y}}, D. 2009, \icarus, 202, 310,
  \dodoi{10.1016/j.icarus.2009.02.033}

\bibitem[{{Morbidelli} {et~al.}(2005){Morbidelli}, {Levison}, {Tsiganis}, \&
  {Gomes}}]{Morbidelli2005}
{Morbidelli}, A., {Levison}, H.~F., {Tsiganis}, K., \& {Gomes}, R. 2005, \nat,
  435, 462, \dodoi{10.1038/nature03540}

\bibitem[{{Mottola} {et~al.}(2011){Mottola}, {Di Martino}, {Erikson},
  {Gonano-Beurer}, {Carbognani}, {Carsenty}, {Hahn}, {Schober}, {Lahulla},
  {Delb{\`o}}, \& {Lagerkvist}}]{Mottola2011}
{Mottola}, S., {Di Martino}, M., {Erikson}, A., {et~al.} 2011, \aj, 141, 170,
  \dodoi{10.1088/0004-6256/141/5/170}

\bibitem[{Napier(2020)}]{Napier_spacerock}
Napier, K. 2020, SpaceRocks, \url{https://github.com/kjnapier/spacerocks}

\bibitem[{{Nesvorny} {et~al.}(2013){Nesvorny}, {Vokrouhlicky}, \&
  {Morbidelli}}]{Nesvorny2013}
{Nesvorny}, D., {Vokrouhlicky}, D., \& {Morbidelli}, A. 2013, in AAS/Division
  for Planetary Sciences Meeting Abstracts, Vol.~45, AAS/Division for Planetary
  Sciences Meeting Abstracts \#45, 508.03

\bibitem[{{Popescu} {et~al.}(2016){Popescu}, {Licandro}, {Morate}, {de
  Le{\'o}n}, {Nedelcu}, {Rebolo}, {McMahon}, {Gonzalez-Solares}, \&
  {Irwin}}]{Popescu2016}
{Popescu}, M., {Licandro}, J., {Morate}, D., {et~al.} 2016, \aap, 591, A115,
  \dodoi{10.1051/0004-6361/201628163}

\bibitem[{{Roig} \& {Nesvorn{\'y}}(2015)}]{Nesvor2015}
{Roig}, F., \& {Nesvorn{\'y}}, D. 2015, \aj, 150, 186,
  \dodoi{10.1088/0004-6256/150/6/186}

\bibitem[{{Roig} {et~al.}(2008){Roig}, {Ribeiro}, \& {Gil-Hutton}}]{Roig2008}
{Roig}, F., {Ribeiro}, A.~O., \& {Gil-Hutton}, R. 2008, \aap, 483, 911,
  \dodoi{10.1051/0004-6361:20079177}

\bibitem[{{Sergeyev} {et~al.}(2022){Sergeyev}, {Carry}, {Onken}, {Devillepoix},
  {Wolf}, \& {Chang}}]{Sergeyev2022}
{Sergeyev}, A.~V., {Carry}, B., {Onken}, C.~A., {et~al.} 2022, \aap, 658, A109,
  \dodoi{10.1051/0004-6361/202142074}

\bibitem[{{Szab{\'o}} {et~al.}(2007){Szab{\'o}}, {Ivezi{\'c}}, {Juri{\'c}}, \&
  {Lupton}}]{Szab2007}
{Szab{\'o}}, G.~M., {Ivezi{\'c}}, {\v{Z}}., {Juri{\'c}}, M., \& {Lupton}, R.
  2007, \mnras, 377, 1393, \dodoi{10.1111/j.1365-2966.2007.11687.x}

\bibitem[{{Uehata} {et~al.}(2022){Uehata}, {Terai}, {Ohtsuki}, \&
  {Yoshida}}]{Uehata2022}
{Uehata}, K., {Terai}, T., {Ohtsuki}, K., \& {Yoshida}, F. 2022, \aj, 163, 213,
  \dodoi{10.3847/1538-3881/ac5b6d}

\bibitem[{Virtanen {et~al.}(2020)Virtanen, Gommers, Oliphant, Haberland, Reddy,
  Cournapeau, Burovski, Peterson, Weckesser, Bright, {van der Walt}, Brett,
  Wilson, Millman, Mayorov, Nelson, Jones, Kern, Larson, Carey, Polat, Feng,
  Moore, {VanderPlas}, Laxalde, Perktold, Cimrman, Henriksen, Quintero, Harris,
  Archibald, Ribeiro, Pedregosa, {van Mulbregt}, \& {SciPy 1.0
  Contributors}}]{2020SciPy-NMeth}
Virtanen, P., Gommers, R., Oliphant, T.~E., {et~al.} 2020, Nature Methods, 17,
  261, \dodoi{10.1038/s41592-019-0686-2}

\bibitem[{{Walsh} {et~al.}(2011){Walsh}, {Morbidelli}, {Raymond}, {O'Brien}, \&
  {Mandell}}]{Walsh2011}
{Walsh}, K.~J., {Morbidelli}, A., {Raymond}, S.~N., {O'Brien}, D.~P., \&
  {Mandell}, A.~M. 2011, \nat, 475, 206, \dodoi{10.1038/nature10201}

\bibitem[{{Wong} \& {Brown}(2015)}]{Wong2015}
{Wong}, I., \& {Brown}, M.~E. 2015, \aj, 150, 174,
  \dodoi{10.1088/0004-6256/150/6/174}

\bibitem[{{Wong} \& {Brown}(2016)}]{Wongbrown2016}
---. 2016, \aj, 152, 90, \dodoi{10.3847/0004-6256/152/4/90}

\bibitem[{{Wong} {et~al.}(2019){Wong}, {Brown}, {Blacksberg}, {Ehlmann}, \&
  {Mahjoub}}]{Wong2019}
{Wong}, I., {Brown}, M.~E., {Blacksberg}, J., {Ehlmann}, B.~L., \& {Mahjoub},
  A. 2019, \aj, 157, 161, \dodoi{10.3847/1538-3881/ab0e00}

\bibitem[{{Wong} {et~al.}(2014){Wong}, {Brown}, \& {Emery}}]{Wong2014}
{Wong}, I., {Brown}, M.~E., \& {Emery}, J.~P. 2014, \aj, 148, 112,
  \dodoi{10.1088/0004-6256/148/6/112}

\bibitem[{{Wong} {et~al.}(2017){Wong}, {Brown}, \& {Emery}}]{Wong2017}
---. 2017, \aj, 154, 104, \dodoi{10.3847/1538-3881/aa8406}

\bibitem[{{Yoshida} \& {Nakamura}(2005)}]{Yoshida2005}
{Yoshida}, F., \& {Nakamura}, T. 2005, \aj, 130, 2900, \dodoi{10.1086/497571}

\bibitem[{{Yoshida} \& {Nakamura}(2008)}]{Yoshida2008}
---. 2008, \pasj, 60, 297, \dodoi{10.1093/pasj/60.2.297}

\bibitem[{{Yoshida} \& {Terai}(2017)}]{Yoshida2017}
{Yoshida}, F., \& {Terai}, T. 2017, \aj, 154, 71,
  \dodoi{10.3847/1538-3881/aa7d03}

\bibitem[{{Yoshida} {et~al.}(2019){Yoshida}, {Terai}, {Ito}, {Ohtsuki},
  {Lykawka}, {Hiroi}, \& {Takato}}]{Yoshida2019}
{Yoshida}, F., {Terai}, T., {Ito}, T., {et~al.} 2019, \planss, 169, 78,
  \dodoi{10.1016/j.pss.2019.02.003}

\end{thebibliography}
\bibliographystyle{aasjournal}


\end{CJK*}
\end{document}